\documentclass{aa}
\usepackage{graphicx}
\usepackage{txfonts}

\begin{document}
\title{
Warm water vapor envelope in Mira variables 
and its effects on the apparent size from the near-infrared 
to the mid-infrared
}

\author{Keiichi Ohnaka}

\offprints{Keiichi Ohnaka, \\ \email{kohnaka@mpifr-bonn.mpg.de}}

\institute{
Max-Planck-Institut f\"{u}r Radioastronomie, 
Auf dem H\"{u}gel 69, D-53121 Bonn, Germany
}

\date{Received / Accepted }

\abstract{
We present a possible interpretation for the increase of the 
angular diameter of the Mira variables $o$~Cet, R~Leo, and $\chi$~Cyg 
from the $K$ band to the 11~\mbox{$\mu$m}\  region revealed by the recent 
interferometric observations using narrow bandpasses where no salient 
spectral feature is present 
(Weiner et al. \cite{weiner03a}, \cite{weiner03b}). 
A simple two-layer model consisting of hot and cool \mbox{H$_2$O}\ layers 
for the warm water vapor envelope, whose
presence in Mira variables was revealed by previous spectroscopic 
observations, can reproduce the angular diameters observed with 
Infrared Spatial Interferometer 
as well as the high-resolution TEXES spectra obtained in the 11~\mbox{$\mu$m}\ 
region.   The warm water vapor layers are optically thick 
in the lines, and therefore, strong absorption due to \mbox{H$_2$O}\ can be 
expected from such a dense water vapor envelope.  
However, the absorption lines are filled in by emission from the
extended part of the envelope, and this results in the 
high-resolution 11~\mbox{$\mu$m}\  spectra which exhibit only weak, 
fine spectral features, masking the spectroscopic evidences of the dense, 
warm water vapor envelope.  On the other hand, the presence of the warm 
water vapor envelope manifests itself as the larger angular diameters 
in the 11~\mbox{$\mu$m}\  region as compared to those measured in the
near-infrared.  
Furthermore, comparison of the visibilities predicted in the near-infrared 
(1.5--3.8~\mbox{$\mu$m}) with observational results available in the
literature demonstrates that 
our two-layer model for the warm water vapor envelope can also reproduce 
the observed near-infrared visibilities and angular diameters, and suggests 
that the wavelength dependence of the angular size of Mira variables 
in the infrared largely reflects the \mbox{H$_2$O}\ opacity.  
The radii of the hot \mbox{H$_2$O}\ layers in the three Mira variables are 
derived to be 1.5--1.7~\mbox{$R_{\star}$}\  with temperatures of 1800--2000~K 
and \mbox{H$_2$O}\  column densities of 
$(1 \mbox{--} 5)  \times 10^{21}$~\mbox{cm$^{-2}$}, 
while the radii of the cool \mbox{H$_2$O}\ layers are derived to be 
2.2--2.5~\mbox{$R_{\star}$}\ 
with temperatures of 1200--1400~K and \mbox{H$_2$O}\  column densities of 
$(1 \mbox{--} 7) \times 10^{21}$~\mbox{cm$^{-2}$}.  
}   

\titlerunning{Warm water vapor envelope in Mira variables}
\authorrunning{K.~Ohnaka}
\maketitle

\section{Introduction}
\label{sect_intro}

An increase of the angular size from the near-infrared toward longer 
wavelengths appears to be a common phenomenon observed in 
various classes of cool luminous stars such as M (super)giants and 
Mira variables.  Mennesson et al. (\cite{mennesson02}) revealed 
that the uniform disk (UD) diameters of semiregular M giants as well as 
Mira variables measured in the \mbox{$L^{\prime}$}\ band are 20--100\% larger 
than those measured in the \mbox{$K^{\prime}$}\ band.  
Recently, Weiner et al. (\cite{weiner00}, \cite{weiner03a}, and 
\cite{weiner03b}, hereafter W00, WHT03a, and WHT03b, respectively) have 
measured the angular size at 11~\mbox{$\mu$m}\ for 
the M supergiants \mbox{$\alpha$~Ori}\
and \mbox{$\alpha$~Her}\ as well as the Mira variables \mbox{$o$~Cet}, 
\mbox{R~Leo}, and \mbox{$\chi$~Cyg}, 
using the Infrared Spatial Interferometer (ISI) with a spectral
bandwidth of 0.17~\mbox{cm$^{-1}$}.  
They found that the uniform disk diameters of the two M supergiants 
measured at 11~\mbox{$\mu$m}\ 
are $\sim$30\% larger than those measured in the $K$ band, 
while the 11~\mbox{$\mu$m}\ uniform disk diameters of the three Mira variables 
are roughly twice as large as those measured in the $K$ band.  

Although these Mira variables exhibit dust emission from the
circumstellar envelope, the observed increase of the 
angular diameter from the near-infrared to the mid-infrared cannot 
simply be attributed to the dust shell, as discussed in detail by 
WHT03a.  The radii of the inner boundary of the circumstellar dust shell 
are estimated to be 2--4~\mbox{$R_{\star}$}\ for \mbox{$o$~Cet}\ and 
\mbox{R~Leo}, and $\sim$20~\mbox{$R_{\star}$}\ for \mbox{$\chi$~Cyg}\ 
(Danchi et al. \cite{danchi94}, Schuller et al. \cite{schuller04}), 
and such extended dust shells 
are completely resolved with the baselines of 20--56~m 
used in the ISI observations by W00, WHT03a, and WHT03b.  
In other words, 
the visibility component resulting from the dust shell is nearly
zero at these baseline lengths, 
and the effect of the dust shell is to lower the total visibility 
by an amount equal to the fraction of flux originating in the dust 
shell in the field of view.  This effect is already taken into account 
in the determination of the uniform disk diameters by W00, WHT03a, and 
WHT03b, and therefore, it is unlikely that the extended dust shell 
causes the 11~\mbox{$\mu$m}\ angular size to appear larger than in the 
near-infrared. 

Instead, the increase of the angular size from the $K$ band 
to the \mbox{$L^{\prime}$}\ band as well as to the 11~\mbox{$\mu$m}\
region can be 
attributed to extended gaseous layers close to the photosphere.  
Mennesson et al. (\cite{mennesson02}) conclude that the observed 
increase of the angular diameter from the \mbox{$K^{\prime}$}\ band to 
the \mbox{$L^{\prime}$}\ 
band can be explained by an optically thin layer ($\tau \approx 0.5$) 
extending to $\sim$3~\mbox{$R_{\star}$}\ 
with a (pseudo)continuous opacity.  Perrin et al. (\cite{perrin04}) 
apply a similar model to interferometric observations of \mbox{$\alpha$~Ori}\ 
and \mbox{$\alpha$~Her}\ obtained in the $K$ and $L$ bands as well as in the 
11~\mbox{$\mu$m}\ region, and conclude that the extended gaseous layer is 
optically thin in the $K$ and $L$ bands ($\tau_K = 0.06$ and 
$\tau_L = 0.026$), but optically thick in the 11~\mbox{$\mu$m}\ region 
($\tau_{11\mu{\rm m}} = 2.33$).  Both Mennesson et al. (\cite{mennesson02}) 
and Perrin et al. (\cite{perrin04}) suspect opacities due to molecular 
species such as \mbox{H$_2$O}, CO, and SiO as the source of the
(pseudo)continuous opacity.  
The presence of such a warm molecular envelope is already confirmed 
by analyses of infrared molecular spectra.  Tsuji (\cite{tsuji78}) 
interpreted infrared spectral data of the M supergiant
$\mu$~Cep and the Mira variable R~Cas in terms of a warm water vapor 
envelope, and Tsuji (\cite{tsuji88}) 
further found evidence of the presence of the warm 
molecular envelope in the high-resolution spectra of M (super)giants. 
More recently, infrared spectra obtained with the Short Wavelength 
Spectrometer (SWS) onboard the Infrared Space Observatory (ISO) 
have provided ample information on physical parameters 
of the warm molecular envelope (e.g., Tsuji et al. \cite{tsuji97}, 
Yamamura et al.~\cite{yamamura99}, Cami et al. \cite{cami00}, 
Matsuura et al.~\cite{matsuura02}).  

However, the high-resolution 11~\mbox{$\mu$m}\ spectra of the M supergiants 
\mbox{$\alpha$~Ori}\ and \mbox{$\alpha$~Her}\ as well as the 
Mira variables \mbox{$o$~Cet}, \mbox{R~Leo}, 
and \mbox{$\chi$~Cyg}\ presented by WHT03a and WHT03b turn out to 
pose difficulty in interpreting the increase of the angular 
diameter in terms of the extended, warm water vapor envelope.  
These 11~\mbox{$\mu$m}\ spectra were obtained using the TEXES instrument
mounted on the Infrared Telescope Facility telescope 
with a spectral resolution of $\sim$10$^5$ (Lacy et al. \cite{lacy02}).  
The high-resolution TEXES spectra of the above M supergiants and 
Mira variables reveal that no salient spectral 
features are present within the bandpasses used in the ISI observations, 
and therefore, these 
spectroscopic observations appear to be in contradiction with the 
interpretation that the dense, warm water vapor envelope is responsible 
for the increase of the angular size from the near-infrared to the 
11~\mbox{$\mu$m}\ region.  

Ohnaka (\cite{ohnaka04}, hereafter Paper~I) 
demonstrates that the warm water vapor envelope 
extending to 1.4--1.5~\mbox{$R_{\star}$}\ with temperatures of 
$\sim$2000~K and \mbox{H$_2$O}\ column densities of the order of 
$10^{20}$~\mbox{cm$^{-2}$}\ can 
reproduce the observed increase of the angular diameter 
from the near-infrared to the 11~\mbox{$\mu$m}\ region in 
\mbox{$\alpha$~Ori}\ and \mbox{$\alpha$~Her}\ and, simultaneously, 
their featureless high-resolution spectra at 11~\mbox{$\mu$m}.  
While significant absorption is expected from such a warm 
water vapor envelope, the absorption lines are filled in by the emission 
of \mbox{H$_2$O}\ lines from the extended part of the envelope, which leads to 
spectra without any conspicuous spectral features as 
observed with the TEXES instrument.  
On the other hand, the emission from the extended water vapor envelope 
causes the apparent size of the objects to be significantly larger than 
the photospheric size, in agreement with the interferometric observations 
with ISI presented by WHT03b.  

If the warm water vapor envelope is present even in early M supergiants 
such as \mbox{$\alpha$~Ori}\ and $\mu$~Cep and causes the stellar 
angular size in the mid-infrared to appear larger than that 
measured in the near-infrared, it can also be the case for Mira 
variables, for which water vapor is known to be present in 
the photosphere and we can expect even denser water vapor 
envelopes than in early M supergiants.  
In the present paper, we attempt to interpret the increase 
of the angular diameter from the near-infrared to the 11~\mbox{$\mu$m}\ region 
as well as the high-resolution 11~\mbox{$\mu$m}\ TEXES spectra observed for 
the Mira variables \mbox{$o$~Cet}, \mbox{R~Leo}, 
and \mbox{$\chi$~Cyg}\ in terms of the warm water vapor envelope.

\section{Modeling of the warm water vapor envelope}

\begin{figure}
\begin{center}
\resizebox{8cm}{!}{\rotatebox{0}{\includegraphics{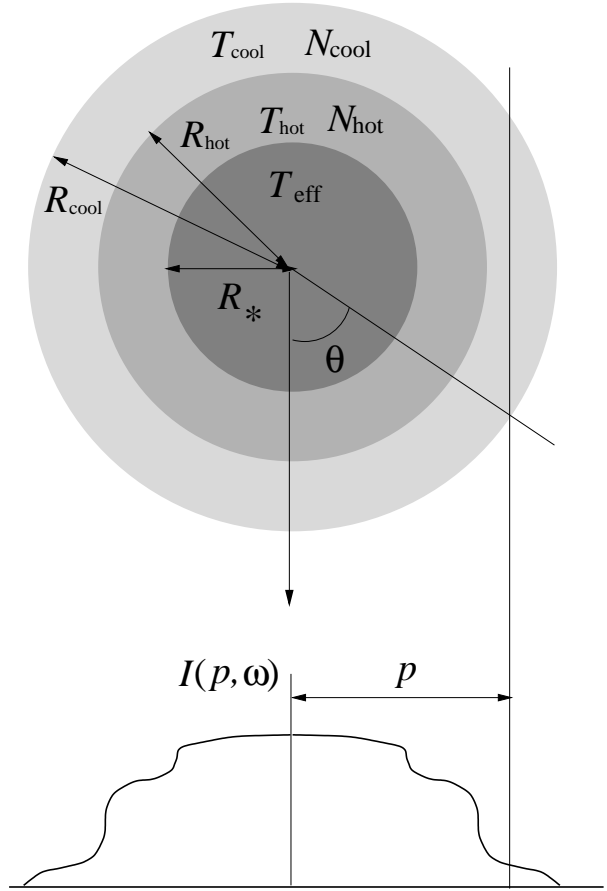}}}
\end{center}
\caption{The model used in the present work.   The photosphere is 
approximated by a blackbody of \mbox{$T_{\rm eff}$}\ with a 
radius \mbox{$R_{\star}$}.  
The column densities in the radial direction 
(\mbox{$N_{\rm hot}$}\ and \mbox{$N_{\rm cool}$}) 
and the temperatures of \mbox{H$_2$O}\ gas (\mbox{$T_{\rm hot}$}\ 
and \mbox{$T_{\rm cool}$}) 
as well as the radii of the water vapor layers (\mbox{$R_{\rm hot}$}\
 and \mbox{$R_{\rm cool}$}) are the input parameters.  
}
\label{model}
\end{figure}

In our model, the central star is surrounded by a warm water vapor 
envelope, which consists of two layers with different temperatures 
\mbox{$T_{\rm hot}$}\ and \mbox{$T_{\rm cool}$}, as depicted in 
Fig.~\ref{model}.  The central 
star is approximated by a blackbody of effective temperature
\mbox{$T_{\rm eff}$} with a radius \mbox{$R_{\star}$}.  The temperature 
and the density are assumed to be constant in these hot and cool 
layers, which extend to 
\mbox{$R_{\rm hot}$}\ and \mbox{$R_{\rm cool}$}, respectively.  
We denote the column density of \mbox{H$_2$O}\ 
in the radial direction in the hot and cool layers as 
\mbox{$N_{\rm hot}$}\ and \mbox{$N_{\rm cool}$}, respectively.  
As very simple it may seem, such a two-layer model turned out to be 
successful in interpreting infrared molecular spectra obtained 
with ISO/SWS.  For example, Yamamura et al. (\cite{yamamura99}) and 
Matsuura et al. (\cite{matsuura02}) could reproduce \mbox{H$_2$O}\ spectra of 
Mira variables obtained with ISO/SWS using a two-layer model, 
although their two-layer model consists 
of a stack of plane-parallel slabs, instead of spherical layers as
adopted in the present work.  

We adopt an effective temperature of 3000~K for the Mira variables 
studied in the present work.  While the effective temperatures of 
Mira variables are known to show temporal variations with phase 
(e.g., van Belle et al. \cite{vanbelle96}, Perrin et al. \cite{perrin99}, 
Woodruff et al. \cite{woodruff04}), the effect of the uncertainty 
of the effective temperature adopted in our model turns out to be 
minor as we show below, 
because the warm water vapor envelope is quite optically thick 
in the wavelength regions discussed in the present work.   

We first calculate the line opacity due to \mbox{H$_2$O}\ assuming a Gaussian 
profile.  We adopt a velocity of 5~\mbox{km s$^{-1}$}\ for the sum of the 
thermal velocity and the micro-turbulent velocity.  The $g\! f$-values 
and the lower excitation potentials of \mbox{H$_2$O}\ lines are calculated 
using the HITEMP line list (Rothman \cite{rothman97}) and the list 
compiled by Partridge \& Schwenke (\cite{partridge97}, hereafter
PS97).  The HITEMP line list is an extension of the HITRAN database 
toward high temperatures expected in stellar atmospheres, and the HITEMP 
\mbox{H$_2$O}\ line list used here is calculated for a temperature of 1000~K.  
As we demonstrate below, the temperature of the warm water vapor envelope 
can be as high as $\sim$2000~K, and therefore, the HITEMP \mbox{H$_2$O}\ line 
list for 1000~K may not be adequate for such a high temperature.  
The PS97 \mbox{H$_2$O}\ line list is generated for a temperature of 4000~K 
and appears to be the most extensive line list of \mbox{H$_2$O}\ up to date.  
We use both line lists for our calculation to study possible effects 
of the difference between the two \mbox{H$_2$O}\ line lists. 

The energy level populations of \mbox{H$_2$O}\ are calculated assuming 
local thermodynamical equilibrium (LTE).  
We examine the validity of LTE in the temperature and density ranges 
relevant for the present work, using the order-of-magnitude estimates 
of collisional and radiative de-excitation rates, as adopted in Paper~I.
The collisional de-excitation rate $C_{\rm ul}$ is given by 
$C_{\rm ul} \sim \! N \sigma_{\rm ul} v_{\rm rel}$, 
where $N$ is the density of the primary collision partner, $\sigma_{\rm ul}$ 
is the cross section, which we approximate with the geometrical 
cross section, and $v_{\rm rel}$ is the 
relative velocity between the collision partner species 
and \mbox{H$_2$O}\ molecules.  
As we will show below, for the hot layer of the warm water vapor
envelope in \mbox{$o$~Cet}, \mbox{R~Leo}, and \mbox{$\chi$~Cyg}, 
the column density of \mbox{H$_2$O}\ 
is found to be of the order of $10^{21}$~\mbox{cm$^{-2}$}\ with temperatures 
of 1800--2000~K, and the geometrical thickness is derived to be 
$\sim$0.5~\mbox{$R_{\star}$}, which translates into 
$1 \times 10^{13}$~cm with a 
stellar radius of 300~\mbox{$R_{\sun}$}\ assumed.  
The number density of \mbox{H$_2$O}\ is 
then estimated to be $1 \times 10^{8}$~\mbox{cm$^{-3}$}.   For temperatures 
and densities found in the hot layer of the warm water vapor envelope, 
H atoms appear to be the primary collision partner of 
\mbox{H$_2$O}\ molecules, 
and the ratio of the number density of H atoms to that of \mbox{H$_2$O}\
molecules expected in chemical equilibrium is approximately $10^{4}$ 
for the relevant temperatures and densities.  
Therefore, the number density of H atoms in the hot layer is estimated to be 
$\sim \! 1 \times 10^{12}$~\mbox{cm$^{-3}$}.  With a geometrical 
cross section $\sigma_{\rm ul}$ of $10^{-15}$~cm$^2$ and a relative 
velocity $v_{\rm rel}$ of 5~\mbox{km s$^{-1}$}\ assumed, 
this number density of 
H atoms leads to a collisional de-excitation rate of 
$\sim \! 500$~s$^{-1}$.  For the cool component of the water vapor
envelope, we derive temperatures of 1200--1400~K, geometrical 
thicknesses of $\sim \! 1.5$~\mbox{$R_{\star}$}, and 
\mbox{H$_2$O}\ column densities of 
the order of $10^{21}$~\mbox{cm$^{-2}$}, which leads to a number density of
\mbox{H$_2$O}\ of $\sim \! 3 \times 10^7$~\mbox{cm$^{-3}$}.  
At these temperatures and densities, 
H$_2$ molecules are the main collision partner of \mbox{H$_2$O}\ molecules, 
and the ratio of the number density of H$_2$ molecules to that of 
\mbox{H$_2$O}\ molecules expected in chemical equilibrium is 
$\sim \! 3 \times 10^3$, which leads to a collisional de-excitation rate 
of $\sim \! 50$~s$^{-1}$.  
On the other hand, the rate of spontaneous emission can be 
estimated from the Einstein coefficients $A_{\rm ul}$.  
For the \mbox{H$_2$O}\ molecule, the ranges of $A_{\rm ul}$ are approximately 
$\sim \! \! 10^{-4}$--$3 \times 10^{2}$~s$^{-1}$, 
$\sim \! \! 10^{-5}$--$3 \times 10^{2}$~s$^{-1}$, and 
$\la 10$~s$^{-1}$ in the 11~\mbox{$\mu$m}\ region, 
$K$ band, and \mbox{$L^{\prime}$}\ band, respectively.  Therefore, 
for these wavelength regions we will discuss below, 
the assumption of LTE is valid for weak and moderately strong \mbox{H$_2$O}\
lines both in the hot and cool layers of the warm water vapor 
envelope, 
while non-LTE effects may not be negligible for very strong lines.  
However, a comprehensive calculation with non-LTE effects 
taken into account is beyond the scope of the present paper, and 
we assume LTE for the \mbox{H$_2$O}\ lines considered here. 

Once the line opacity has been obtained, the monochromatic intensity 
profile as well as the spectrum (emergent flux) 
can be calculated at an appropriate wavenumber interval, 
as described in Paper~I.  
In order to compare with the observed TEXES spectra of the three 
Mira variables, 
synthetic spectra are spectrally convolved with a Gaussian profile which 
represents the effects of the spectral resolution of the instrument 
as well as of the macro-turbulent velocity in the atmosphere of the Mira
variables.  The spectral resolution of the TEXES instrument is $10^5$, 
which translates into an instrumental broadening of 3~\mbox{km s$^{-1}$}.  
The macro-turbulent velocity in the atmosphere of Mira variables is not 
well known.  For non-Mira M giants, however, analyses of high-resolution 
spectra are available, and they may be used as a guide to estimate 
the macro-turbulent velocity in the atmosphere of Mira variables.  
Tsuji (\cite{tsuji86}) analyzed the CO first-overtone bands of 
non-Mira K and M giants and obtained macro-turbulent velocities 
ranging from $\sim$1~\mbox{km s$^{-1}$}\ to $\sim$4~\mbox{km s$^{-1}$}. 
In the present work, 
we tentatively assume a macro-turbulent velocity of 3~\mbox{km s$^{-1}$}\ for 
the three Miras, and the FWHM of the Gaussian profile to represent 
the spectral resolution of 
the instrument and the macro-turbulent velocity is calculated as 
$\sqrt{3^2 + 3^2} = 4.24$~\mbox{km s$^{-1}$}.   
The spectrum convolved with this Gaussian profile 
and normalized, $F_{\omega}$, is then diluted as follows, taking the
effect of 
continuous dust emission from the circumstellar dust shell into account: 
\begin{equation}
 F_{\omega}^{\rm diluted} = (1 - f_{\rm dust})F_{\omega} + f_{\rm dust} \, ,
\end{equation}
where $F_{\omega}^{\rm diluted}$ is the final spectrum and $f_{\rm dust}$ 
is the fraction of the flux contribution of the circumstellar dust 
shell.   

We calculate the monochromatic visibility from the monochromatic 
intensity profile at each wavenumber 
using the Hankel transform, and then the monochromatic visibility 
is spectrally convolved with an appropriate response function 
which represents the spectral resolution of the interferometric
observation at issue.  For the ISI observations we will discuss below, 
WHT03a and WHT03b used narrow bandpasses with a width of 
0.17~\mbox{cm$^{-1}$}. 
Therefore, the spectral response function is assumed to be a top-hat 
function with a width of 0.17~\mbox{cm$^{-1}$}.  

We compare synthetic spectra and 
uniform disk diameters with the TEXES spectra as well as the 11~\mbox{$\mu$m}\ 
uniform disk diameters obtained by WHT03a and 
WHT03b, searching for a parameter set which can best reproduce these 
observational results.  The ranges of the input parameters are as 
follows: \mbox{$T_{\rm hot}$}\ (K) = 1600 ... 2000 with 
$\Delta \mbox{$T_{\rm hot}$} = 100$~K, 
\mbox{$T_{\rm cool}$}\ (K) = 1000 ... 1400 with 
$\Delta \mbox{$T_{\rm cool}$} = 100$~K, 
\mbox{$R_{\rm hot}$}\ (\mbox{$R_{\star}$}) = 1.3, 1.5, 1.7, and 2.0, 
and \mbox{$R_{\rm cool}$}\ (\mbox{$R_{\star}$}) = 2.0, 2.2, 2.5, 2.7, 
and 3.0 with a condition $\mbox{$R_{\rm hot}$} < \mbox{$R_{\rm cool}$}$.  
The grid for \mbox{$N_{\rm hot}$}\ (\mbox{cm$^{-2}$} ) and 
\mbox{$N_{\rm cool}$}\ (\mbox{cm$^{-2}$} ) is 
$10^{19}$, $10^{20}$, $5 \times 10^{20}$, 
$10^{21}$, $3 \times 10^{21}$, $5 \times 10^{21}$, $7 \times 10^{21}$, 
and $10^{22}$.  
The uncertainties of the derived temperatures and radii of the 
\mbox{H$_2$O}\ layers are estimated to be $\pm 100$~K and 
$\pm 0.2$~\mbox{$R_{\star}$}, 
respectively, while the uncertainties of the \mbox{H$_2$O}\ column densities 
are approximately a factor of 2.  
The adoption of an effective temperature of 2800~K, instead of 3000~K, 
leads to no noticeable changes in the spectra and visibilities discussed 
in the present work. 
In the next section, we discuss comparison between observation and 
model for each object.

\section{Comparison with the observed spectra and angular diameters}

\subsection{\mbox{$o$~Cet}}
\label{sect_omicet}

\subsubsection{11~\mbox{$\mu$m}\ spectrum}

\begin{figure}
\begin{center}
\resizebox{8.5cm}{!}{\rotatebox{-90}{\includegraphics{1207f2.ps}}}
\end{center}
\caption{Spectra of \mbox{$o$~Cet}\ in the 11~\mbox{$\mu$m}\ region 
at phase 0.36. 
The red dots represent the observed spectrum of \mbox{$o$~Cet}\ presented in 
WHT03a, while 
the green and blue solid lines represent the calculated spectra 
using the HITEMP database and the PS97 line list, respectively.  
The parameters of the best-fit model for \mbox{$o$~Cet}\ are 
\mbox{$T_{\rm hot}$}\ = 1800~K, 
\mbox{$R_{\rm hot}$}\ = 1.5~\mbox{$R_{\star}$}, \mbox{$N_{\rm hot}$}\  
= $5 \times 10^{21}$~\mbox{cm$^{-2}$}, \mbox{$T_{\rm cool}$}\ = 1400~K, 
\mbox{$R_{\rm cool}$}\ = 2.2~\mbox{$R_{\star}$}, 
and \mbox{$N_{\rm cool}$}\ = $1 \times 10^{21}$~\mbox{cm$^{-2}$}.  
The synthetic spectra are convolved with a Gaussian with a FWHM of 
0.013~\mbox{cm$^{-1}$}\ to account for the effects of the instrument as well 
as of the macro-turbulent velocity, and are redshifted by 
83~\mbox{km s$^{-1}$}\ to match the observation.  
The dashed lines represent the bandpasses used in the ISI observations. 
The positions of the \mbox{H$_2$O}\ lines whose 
intensity at 2000~K is stronger than 
$1 \times 10^{-24}$~cm molecule$^{-1}$ are marked with upper ticks 
(PS97 line list) and lower ticks (HITEMP database).  
These line positions are also redshifted by 83~\mbox{km s$^{-1}$}\ with 
respect to the rest wavenumber. 
}
\label{omiCetSp11mu}
\end{figure}

\begin{figure}
\begin{center}
\resizebox{8.5cm}{!}{\rotatebox{-90}{\includegraphics{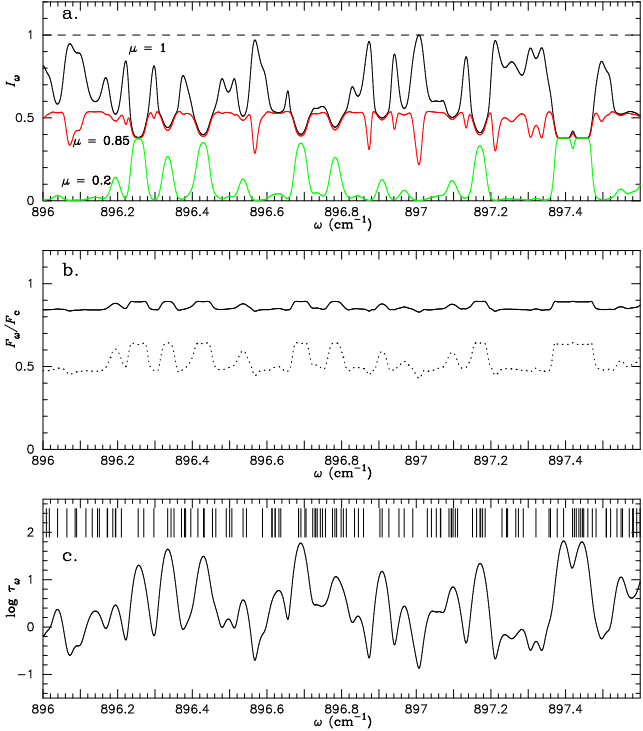}}}
\end{center}
\caption{{\bf a}: 11~\mbox{$\mu$m}\ spectra predicted from the best-fit 
model for \mbox{$o$~Cet}\ at the stellar disk center 
($\mu = 1$, top spectrum), near the limb of the warm \mbox{H$_2$O}\ 
envelope ($\mu = 0.2$, bottom spectrum), and between the disk center 
and the limb ($\mu = 0.85$, middle spectrum).  
The continuum level is shown with the dashed line, and the HITEMP 
database is used in the calculation.  
The flux contribution of dust emission is not included in the spectra shown.  
{\bf b}: The dotted line represents the spectrum 
(emergent flux) expected for the whole object (stellar disk + 
warm water vapor envelope) without the dilution effect due to  
dust emission.  The solid line represents the final spectrum with a 
flux contribution of dust emission of 70\% taken into account.  
{\bf c}: Optical depth in the radial direction due to \mbox{H$_2$O}\ lines 
in the same spectral region.  The positions of the \mbox{H$_2$O}\ lines are 
marked with ticks.  
The spectra and the flux shown in {\bf a} and {\bf b} 
as well as the optical depth shown in {\bf c} are 
redshifted by 83~\mbox{km s$^{-1}$}\ with respect to the rest wavenumber to 
match the observed spectrum shown in Fig.~\ref{omiCetSp11mu}, 
but {\em not} convolved with the Gaussian representing the 
instrumental effect and the macro-turbulent velocity. 
}
\label{omiCetIntensSp11mu}
\end{figure}

For \mbox{$o$~Cet}, 
we find that the 11~\mbox{$\mu$m}\ spectrum observed at phase
0.36, which is presented in WHT03a, 
as well as the 11~\mbox{$\mu$m}\ uniform disk diameters obtained by W00 
and WHT03a can be 
best reproduced by a model with \mbox{$T_{\rm hot}$}\ = 1800~K, 
\mbox{$R_{\rm hot}$}\ = 1.5~\mbox{$R_{\star}$}, 
\mbox{$N_{\rm hot}$}\ = $5 \times 10^{21}$~\mbox{cm$^{-2}$}, 
\mbox{$T_{\rm cool}$}\ = 1400~K, \mbox{$R_{\rm cool}$}\ = 
2.2~\mbox{$R_{\star}$}, 
and \mbox{$N_{\rm cool}$}\ = $1 \times 10^{21}$~\mbox{cm$^{-2}$}.  
Figure~\ref{omiCetSp11mu} shows the synthetic spectra in the 11~\mbox{$\mu$m}\ 
region predicted by the best-fit model for \mbox{$o$~Cet}, together with 
the TEXES spectrum obtained at phase 0.36, which was read off Fig.~2 
in WHT03a.  
The synthetic spectrum calculated with the HITEMP \mbox{H$_2$O}\ line list 
is represented by the green solid line, while that calculated with 
the PS97 \mbox{H$_2$O}\ line list is represented by the blue solid line.  
The synthetic spectra are redshifted by 83~\mbox{km s$^{-1}$}\ to match the 
observation (see WHT03a). 
WHT03a derived the fraction of the flux contribution of the stellar disk  
of \mbox{$o$~Cet}\ ranging from 0.24 to 0.47 at 11~\mbox{$\mu$m}, 
which corresponds to the fraction of the flux contribution of the 
dust shell between 0.76 and 0.53.  We adopted 
$f_{\rm dust} = 0.7$ in our calculation of the 11~\mbox{$\mu$m}\ region for 
\mbox{$o$~Cet}.  
While the synthetic spectrum calculated with the PS97 
line list shows \mbox{H$_2$O}\ features which are not seen in the spectrum 
calculated with the HITEMP database, the discrepancy between the two 
synthetic spectra is not very serious or systematic for 
the temperatures and densities relevant for the present calculation, 
given the uncertainties of the line positions of both line lists.  
For example, Ryde et al. (\cite{ryde02}) 
found that the uncertainty of the positions of \mbox{H$_2$O}\ pure 
rotation lines around 12~\mbox{$\mu$m}\ predicted by PS97 is 
$\sim$0.05~\mbox{cm$^{-1}$}.  On the other hand, the uncertainty of 
\mbox{H$_2$O}\ line positions in the HITEMP database is between
$\sim$0.1 and $\sim$1.0~\mbox{cm$^{-1}$}.  
Figure~\ref{omiCetSp11mu} shows that the two synthetic spectra can 
reproduce the approximate depths and heights of the observed fine spectral 
features and, in particular,  
the observed spectra free from any conspicuous spectral features 
in the bandpasses at around $\sim$896.83~\mbox{cm$^{-1}$}\ and 
$\sim$895.1~\mbox{cm$^{-1}$}\ (dashed lines) 
used for the ISI observations.  
In these bandpasses, a number of \mbox{H$_2$O}\ 
lines are present, as marked by the ticks in Fig.~\ref{omiCetSp11mu}.  
The continuum-like appearance of the spectra is a 
result of the filling-in of the absorption lines by emission from 
the extended part of the warm water vapor envelope, as we will 
discuss below.   
On the other hand, the match for the individual line profiles is not 
very good, which can partially be attributed to the uncertainties of 
the line positions predicted by both line lists, but it is also due 
to the simplicity of our model, which cannot account for complicated 
physical and chemical processes taking place in the outer atmosphere 
of Mira variables.  For example, the strong \mbox{H$_2$O}\ lines at 897.40, 
897.45, 902.04, and 918.57~\mbox{cm$^{-1}$}\ show inverse P-Cygni profiles, 
which suggest an inflow with a velocity of 11~\mbox{km s$^{-1}$}.  
Such a dynamical 
process is not taken into account in our model, and therefore, these 
line profiles cannot be well reproduced by our two-layer model.  

The continuum-like appearance of the spectrum of \mbox{$o$~Cet}\ 
in the 11~\mbox{$\mu$m}\ region can be explained by the filling-in effect 
due to \mbox{H$_2$O}\ line emission from the extended part of the warm 
water vapor envelope.  This effect is illustrated in 
Fig.~\ref{omiCetIntensSp11mu}, where the spectra expected at the disk 
center ($\mu = 1$, where $\mu \equiv \cos \theta$ with $\theta$ defined 
in Fig.~\ref{model}), at a position between the disk center and the 
limb ($\mu = 0.85$), and near the limb ($\mu = 0.2$) are shown in 
the top panel.  
The bottom panel of Fig.~\ref{omiCetIntensSp11mu} shows the optical 
depth of the water vapor envelope, and the figure reveals that 
the water vapor envelope is optically thick 
($\tau_{\omega} \approx 1$--60).  
These spectra as well as the optical depth are calculated 
with the HITEMP database.  
The \mbox{H$_2$O}\ lines appear as strong absorption in the spectrum 
at the disk center, as the top spectrum in 
Fig.~\ref{omiCetIntensSp11mu}a shows.  On the other hand, the 
spectrum near the limb (bottom spectrum in Fig.~\ref{omiCetIntensSp11mu}a) 
shows emission at the positions of the \mbox{H$_2$O}\ lines, 
and these emission 
lines originate in the cooler \mbox{H$_2$O}\ layer.  The spectrum for a line 
of sight between the disk center and the limb (middle spectrum in 
Fig.~\ref{omiCetIntensSp11mu}a) is dominated by 
the \mbox{H$_2$O}\ emission lines originating in the hotter layer overlaid by 
the \mbox{H$_2$O}\ absorption lines originating in the cooler layer.  
Since the spectrum (emergent flux) of the object is obtained by integrating 
the intensity over the area that the object projects onto the 
plane of the sky, the absorption due to \mbox{H$_2$O}\ lines is significantly 
filled in by emission from the limb, resulting in the spectrum showing 
only moderately strong emission features, as plotted with the dotted line in 
Fig.~\ref{omiCetIntensSp11mu}b.  These emission features are further 
weakened by dilution due to continuous dust emission, as the final 
spectrum (solid line) in Fig.~\ref{omiCetIntensSp11mu}b shows.  

The parameters derived above are similar to those derived by 
Yamamura et al. (\cite{yamamura99}) for \mbox{$o$~Cet}\ at phase 0.99: 
\mbox{$T_{\rm hot}$}\ = 2000~K, 
\mbox{$R_{\rm hot}$}\ = 2.0~\mbox{$R_{\star}$}, 
\mbox{$N_{\rm hot}$}\ = $3.0 \times 10^{21}$~\mbox{cm$^{-2}$}, 
\mbox{$T_{\rm cool}$}\ = 1400~K, 
\mbox{$R_{\rm cool}$}\ = 2.3~\mbox{$R_{\star}$}, 
and \mbox{$N_{\rm cool}$}\ = $3.0 \times 10^{20}$~\mbox{cm$^{-2}$}. 
However, it is not straightforward to discuss the phase dependence 
of these physical properties, not only because of the uncertainties 
of the derived parameters, but also because of the phase dependence 
of the stellar continuum radius.  The analysis by Woodruff et al. 
(\cite{woodruff04}) demonstrates that the stellar radius of \mbox{$o$~Cet}\ 
in the 1.04~\mbox{$\mu$m}\  continuum predicted by dynamical model atmospheres 
shows a sinusoidal variation from phase 0.0 to 0.5 with a maximum 
at around phase 0.25, while the relatively large errors of the 
observationally derived continuum radii resulting from the uncertainty 
of the distance prevented the authors from confirming this theoretical 
prediction.  Interferometric observations with spectral resolution 
high enough to resolve molecular bands would be useful for obtaining 
reliable stellar continuum diameters and studying temporal variations 
of physical properties of the warm water vapor envelope.  

\subsubsection{11~\mbox{$\mu$m}\ angular diameter}

\begin{figure}
\begin{center}
\resizebox{8.0cm}{!}{\rotatebox{-90}{\includegraphics{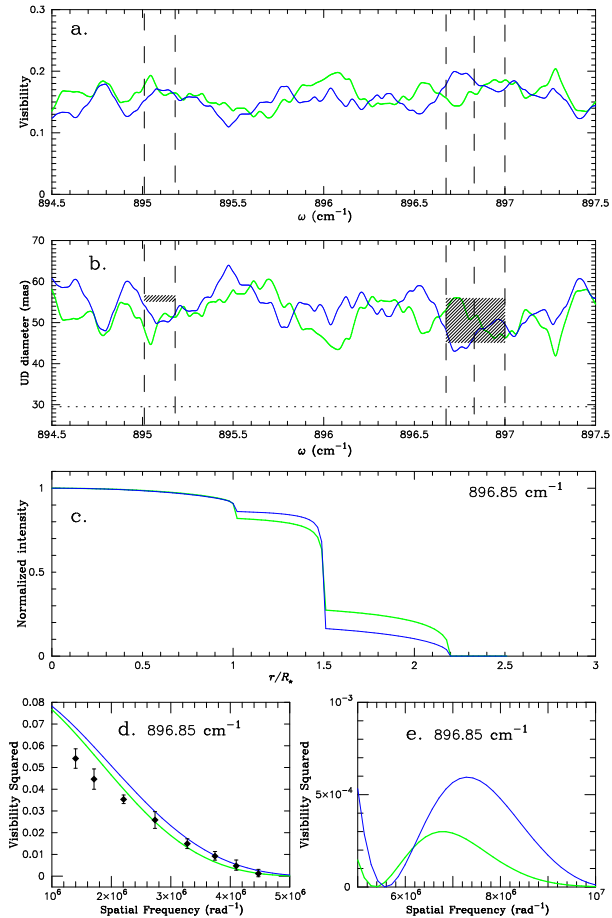}}}
\end{center}
\caption{
{\bf a:} Visibilities calculated from the best-fit model for 
\mbox{$o$~Cet}\ in the 11~\mbox{$\mu$m}\ region for a projected baseline 
length of 30~m. In all the panels, 
the green solid lines represent results calculated with the HITEMP 
database, while the blue solid lines represent those calculated with 
the PS97 list.   
{\bf b:} Uniform disk diameters calculated for a projected baseline 
length of 30~m. 
In {\bf a} and {\bf b}, the bandpasses used by WHT03a are marked 
with the dashed lines.  
The bandpasses used for the observations of \mbox{$o$~Cet}\ at 
around 896.83~\mbox{cm$^{-1}$}\ lie between those shown (see WHT03a).  
The ranges of the uniform disk diameters measured in these bandpasses 
are shown as the hatched regions.  
The dotted line represents the photospheric angular diameter 
adopted in the calculation. 
A flux contribution of 70\% from the dust shell is included  
in the calculations of the model visibilities, but not in 
the computation of the uniform disk diameters.  
Both plots ({\bf a} and {\bf b}) are redshifted by 0.248~\mbox{cm$^{-1}$}\ 
with respect to the rest 
wavenumber, which corresponds to a radial velocity of 83~\mbox{km s$^{-1}$}, 
to match the observed spectrum shown in Fig.~\ref{omiCetSp11mu}. 
{\bf c:} The normalized intensity profile at 896.85~\mbox{cm$^{-1}$}, 
which is spectrally convolved with the same top-hat function as used 
in the calculation of the spectrally convolved visibility shown in {\bf a}.  
{\bf d:} The spectrally convolved squared visibilities at 
896.85~\mbox{cm$^{-1}$}, plotted together with the visibility points observed 
by WHT03a (filled diamonds).  
{\bf e:} The same spectrally convolved visibility squared as in {\bf d}, 
but at higher spatial frequencies. 
}
\label{omiCetVis11mu}
\end{figure}

We now calculate the angular diameter of \mbox{$o$~Cet}\ predicted from the 
above model for the warm water vapor envelope.
In order to compare with the observed uniform disk diameters, 
it is necessary to estimate the angular size of the star, which 
corresponds to \mbox{$R_{\star}$}.  
Woodruff et al. (\cite{woodruff04}) analyzed 
the $K$-band interferometric data of \mbox{$o$~Cet}\ obtained with VLTI/VINCI 
at phases from 0.13 to 0.40.  
They derived the Rosseland radius as well as the continuum radius at 
1.04~\mbox{$\mu$m}\ at each phase by fitting the observed visibilities with 
theoretical ones predicted by dynamical model atmospheres, and 
found that the continuum diameter at 1.04~\mbox{$\mu$m}\ at phase 0.40 is
340~\mbox{$R_{\sun}$}, which corresponds to 29.5~mas (see Fig.~7 in 
Woodruff et al. \cite{woodruff04}).  In the present calculation, we
adopt this value as the diameter of the stellar disk.  
Figures~\ref{omiCetVis11mu}a and \ref{omiCetVis11mu}b show the predicted 
visibilities and uniform disk diameters, using the HITEMP database 
and the PS97 line list.  
The visibilities as well as the uniform disk 
diameters shown in the figure are calculated for a projected baseline 
length of 30~m, which is approximately the mean of the baseline lengths 
used by W00, WHT03a, and WHT03b.  
The presence of the extended dust shell lowers the visibility by an 
amount equal to the fraction of the flux contribution of the 
dust shell, and therefore, the visibilities resulting from the stellar disk 
and the warm \mbox{H$_2$O}\ envelope are lowered by a factor of 0.3 to 
account for the flux contribution of the dust shell in this 
wavelength region.  Note, however, that the uniform disk 
diameters are computed from the visibilities excluding the dust shell, 
because the effect of the presence of the dust shell is already 
taken into account in the determination of the uniform disk diameters 
by W00, WHT03a, and WHT03b.  The bandpasses used in the ISI 
observations are marked with the dashed lines, and the ranges of the 
uniform disk diameters observed within these bandpasses are shown as 
the hatched regions.  

Figure~\ref{omiCetVis11mu}b demonstrates that 
the predicted uniform disk diameters in the 
bandpasses centered around 896.83~\mbox{cm$^{-1}$}\ are in agreement with 
those observed.  The ISI observations within these bandpasses were 
carried out at various phases between 0.92 and 0.36, and therefore, 
it should be kept in mind that the range of the observed angular 
diameters shown in the figure is affected by a variation of the 
angular diameter with phase.   
The uniform disk diameters predicted within the 
bandpass centered at 895.1~\mbox{cm$^{-1}$}\ are systematically lower than 
the observed value of $55.88 \pm 0.74$~mas.  However, 
the ISI observation for this bandpass was carried out 
at phase 0.26, while the high-resolution TEXES spectrum of \mbox{$o$~Cet}, 
which we used for deriving the parameters of the water vapor envelope, 
was obtained at phase 0.36.  Furthermore, WHT03a note that the uniform 
disk diameter of \mbox{$o$~Cet}\ measured at 
11.1419~\mbox{$\mu$m}\ increased by 
11\% ($\sim$6~mas) from 2000 to 2001, which they attribute to the 
non-periodic variation in the angular size and/or asymmetries.  
Therefore, the difference between the predicted angular diameters 
within the bandpass centered at 895.1~\mbox{cm$^{-1}$}\ and that observed 
can be attributed to the mismatch in phase as well as this non-periodic 
variation and/or asymmetries of the stellar disk and the warm water
vapor envelope.  

Figure~\ref{omiCetVis11mu}c shows the intensity profiles predicted 
at 896.85~\mbox{cm$^{-1}$}, using the HITEMP database (green solid line) and 
the PS97 line list (blue solid line).  The spectrally convolved 
squared visibilities at this wavenumber are plotted as a function of 
spatial frequency for both cases with the HITEMP database and the PS97 
line list in Fig.~\ref{omiCetVis11mu}d, together with the observed 
values for \mbox{$o$~Cet}, 
which were read off Fig.~1 of WHT03a.  The figure reveals that 
the predicted squared visibilities are in agreement with those observed 
at spatial frequencies higher than $\sim \! 2.6 \times 10^6$~rad$^{-1}$, 
while the agreement is slightly poorer at lower spatial frequencies.  
However, this small discrepancy can be due to the variation of the observed 
angular diameter mentioned above, and therefore, is not regarded as 
serious disagreement.  Contemporaneous spectroscopic and
interferometric observations would be desirable for further 
constraining our model for the warm water vapor envelope.  
Figure~\ref{omiCetVis11mu}e shows the predicted squared visibilities 
at higher spatial frequencies, which correspond to baselines as long as 
$\sim$100~m.  Mid-infrared interferometric observations with such 
long baselines would also be useful for examining our model further.  

\begin{figure}
\begin{center}
\resizebox{8.5cm}{!}{\rotatebox{0}{\includegraphics{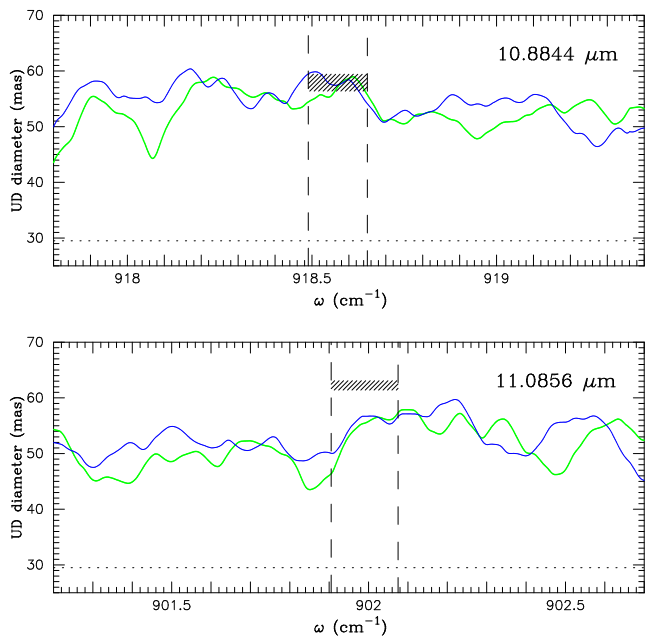}}}
\end{center}
\caption{Uniform disk diameters around 10.8844~\mbox{$\mu$m}\ (upper panel) 
and 11.0856~\mbox{$\mu$m}\ (lower panel) 
calculated from the best-fit model for \mbox{$o$~Cet}\ for a projected 
baseline length of 30~m.  
In both panels, the green solid lines represent results calculated 
with the HITEMP database, while the blue solid lines represent those 
calculated with PS97 list.   
Both plots are redshifted by 0.248~\mbox{cm$^{-1}$}\ in wavenumber, 
which corresponds to a radial velocity of 83~\mbox{km s$^{-1}$}, 
to match the observed spectrum shown in Fig.~\ref{omiCetSp11mu}. 
The bandpasses used by WHT03a are marked with the dashed lines, 
and the ranges of the angular diameters measured by WHT03a are represented 
as the hatched regions.  
The dotted lines represent the photospheric angular diameter 
adopted in the calculation. 
}
\label{omiCetUD11mu}
\end{figure}

We perform the same calculation for the two bandpasses in the regions 
around 10.8844~\mbox{$\mu$m}\ and 11.0856~\mbox{$\mu$m}.  
Within the bandpasses in these wavelength regions, 
strong emission and absorption features due to \mbox{H$_2$O}\ are observed, 
as shown in the upper two panels of Fig.~\ref{omiCetSp11mu}.  
Figure~\ref{omiCetUD11mu} shows the uniform disk diameters calculated 
with the HITEMP database and the PS97 line list in these wavelength 
regions.  The figure demonstrates that 
the observed uniform disk diameter within the bandpass at
918.56~\mbox{cm$^{-1}$}\ can be well reproduced by our model, 
while the uniform disk diameters 
predicted within the bandpass at 902.0~\mbox{cm$^{-1}$}\ are lower than the 
observed value.  It should be noted, however, that the ISI observations 
for these bandpasses were carried out at phases different from the phase 
0.36 at which the TEXES spectrum was obtained: phase 0.95 for the 
bandpass at 918.56~\mbox{cm$^{-1}$}\  and phase 0.20 for the bandpass at 
902.0~\mbox{cm$^{-1}$}.  This mismatch in phase as well as the systematic 
increase of the angular size from 2000 to 2001 mentioned above 
introduces ambiguities in comparing the predicted and observed 
uniform disk diameters.  Given these ambiguities, the agreement between 
the predicted and observed uniform disk diameters within the two 
bandpasses containing the strong \mbox{H$_2$O}\ features can also be 
regarded as fair.  

\subsubsection{Near-infrared angular diameter}

\begin{figure}
\begin{center}
\resizebox{8.5cm}{!}{\rotatebox{-90}{\includegraphics{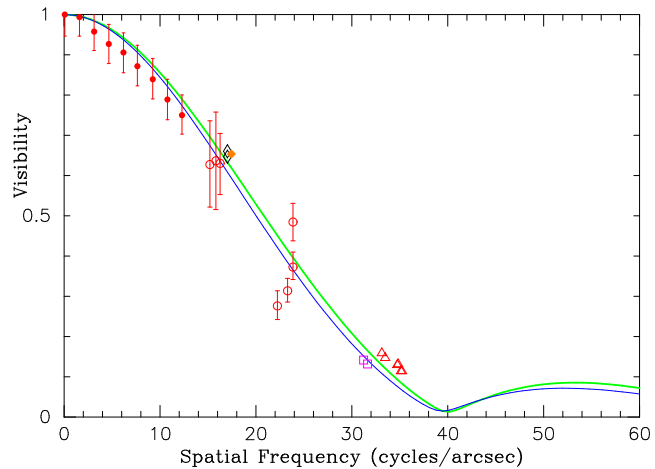}}}
\end{center}
\caption{Comparison between the $K$-band visibilities observed for 
\mbox{$o$~Cet}\ and those predicted by the best-fit model for this object.  
The green solid line represents the visibility calculated with the 
HITEMP database, while the blue solid line represents that 
calculated with PS97 list.   The observed visibilities include 
the results obtained by Ridgway et al. (\cite{ridgway92}) 
(phase 0.23--0.36, red open circles), 
the speckle data obtained at 0.26 by 
Woodruff et al. (\cite{woodruff04}) (red filled circles), 
and the VLTI/VINCI data obtained at phase 0.26 (red open triangles), 
0.40 (purple open squares), 1.40 (black open diamonds), and 1.47 
(orange filled diamonds).  
}
\label{omiCetVisK}
\end{figure}

We have shown that the high-resolution 11~\mbox{$\mu$m}\  spectrum of 
\mbox{$o$~Cet}\ and its 11~\mbox{$\mu$m}\ angular diameter almost twice as large 
as those measured in the near-infrared can be reproduced by our model 
for the warm water vapor envelope.  
However, if the effect of the 
warm water vapor envelope on the near-infrared angular diameter is 
as large as that on the 11~\mbox{$\mu$m}\ angular diameter, our model may 
not reproduce the observed near-infrared visibilities and angular 
diameters of \mbox{$o$~Cet}.  
If it is the case, our model for the warm water vapor envelope 
as well as the photospheric angular diameter of 29.5~mas that we assumed 
based on the $K$-band VINCI observations may not be justified.  
Therefore, we calculate the visibility as well as the angular 
diameter of \mbox{$o$~Cet}\ in the near-infrared using the same best-fit model 
for \mbox{$o$~Cet}\ and compare with the observed values.  

Although a number of near-infrared interferometric observations 
of \mbox{$o$~Cet}\ are presented in the literature, only a limited amount of 
data are available up to now for phases near 0.36.  
Given uncertainties in the determination of a zero point of pulsational 
phase, we can compare with data obtained at phases of approximately 
$0.36 \pm 0.1$.   Ridgway et al. (\cite{ridgway92}) 
obtained $K$-band visibilities between phase 0.23 and 0.36 with baseline 
lengths of 8 and 12~m.  
The $K$-band interferometric data of \mbox{$o$~Cet}\ presented by 
Woodruff et al. (\cite{woodruff04}) include data obtained at phases 
0.26, 0.4, 1.4, and 1.47 (phases larger than 1 mean the next pulsation 
cycle), together with speckle interferometric data obtained with 
the 6~m telescope at 
the Special Astrophysical Observatory in Russia.  
The data obtained by Ridgway et al. (\cite{ridgway92}) as well as 
by Woodruff et al. (\cite{woodruff04}) were acquired with a broadband 
$K$ filter, 
which means that the derived apparent size is affected by 
molecular spectral features originating in the warm molecular envelope, 
in particular, \mbox{H$_2$O}\ lines located at the shorter and longer edges of 
the $K$ band.   We calculate 
the uniform disk diameter in the $K$ band in almost the same manner as in 
the 11~\mbox{$\mu$m}\ region.  The only difference is that we calculate 
the monochromatic visibility squared from the monochromatic intensity 
profile, and this monochromatic visibility squared is spectrally 
convolved for comparison with the above $K$-band visibility measurements. 
We include only \mbox{H$_2$O}\ lines in the calculation, 
and the $K$-band filter 
is approximated with a top-hat function centered at 2.2~\mbox{$\mu$m}\ with 
$\Delta \lambda$ = 0.4~\mbox{$\mu$m}.  
The $K$-band speckle visibility obtained at phase 0.26 shows no evidence of 
an extended dust shell, whose presence would result in a steep drop 
of visibility at low spatial frequencies (see Fig.~2 in Woodruff et al. 
\cite{woodruff04}).  While a possible appearance of significant dust 
emission in the $K$ band at phases other than 0.26 cannot be ruled out, 
it seems to be unlikely that dust emission has a noticeable effect on the 
$K$-band visibility at phase 0.36.  
Therefore, no flux contribution from the dust shell 
is included in the calculation of visibilities in this wavelength region.  
Figure~\ref{omiCetVisK} shows a comparison between the observed $K$-band 
visibilities mentioned above and those predicted by the best-fit model 
for \mbox{$o$~Cet}.  The figure illustrates that the observed visibilities 
are well reproduced by our model.  

We also compare the predicted $K$-band uniform disk diameters with 
that observationally derived.  From the predicted visibilities shown in 
Fig.~\ref{omiCetVisK}, the uniform disk diameter 
is derived for a projected baseline length of 15~m, which is
roughly the mean of the baselines used in the observations 
presented by Woodruff et al. (\cite{woodruff04}).  The $K$-band 
uniform disk diameters predicted from the best-fit model for \mbox{$o$~Cet}\ 
are 34.1~mas with the HITEMP database and 34.9~mas with the PS97 line list.  
These predicted values are in agreement with the uniform 
disk diameter of $33.27 \pm 0.33$~mas at phase 0.40 obtained 
by Woodruff et al (\cite{woodruff04}), and this agreement 
lends support to our model for the warm water vapor envelope as well as 
the photospheric angular diameter adopted in the present calculation.  

\begin{figure}
\begin{center}
\resizebox{8.5cm}{!}{\rotatebox{-90}{\includegraphics{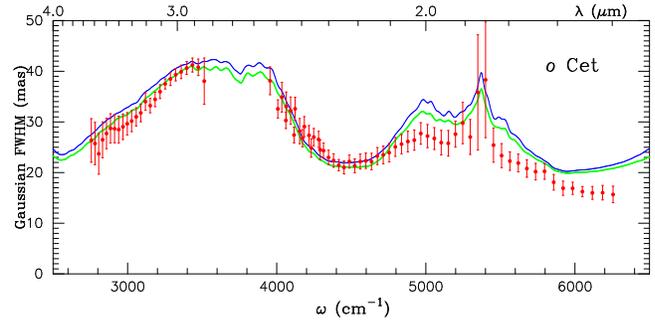}}}
\end{center}
\caption{Wavelength dependence of the angular size (Gaussian FWHM) 
predicted from the best-fit model for \mbox{$o$~Cet}\ for a projected baseline 
length of 10~m.  The red dots represent the angular size of \mbox{$o$~Cet}\ 
observed by Ireland et al. (\cite{ireland04}).  
The green solid line represents the result 
obtained with the HITEMP database, while the blue solid line represents 
that obtained with the PS97 line list.  
No flux contribution from the dust shell is 
included in the calculation shown.  
}
\label{omiCetUD_HKL}
\end{figure}

Ireland et al. (\cite{ireland04}) measured the angular size of 
\mbox{$o$~Cet}\ in the near-infrared between 1.2 and 3.6~\mbox{$\mu$m}\ 
by aperture masking with the Keck telescope.  The wavelength dependence 
of the angular size in the near-infrared appears to coincide with 
the wavelength dependence of the opacity of \mbox{H$_2$O}, which suggests that 
the observed near-infrared angular size may also be explained by 
our model for the warm water vapor envelope.  
We calculate the angular size, which is the Gaussian 
FWHM in this case, using the best-fit model for \mbox{$o$~Cet}, with only 
\mbox{H$_2$O}\ lines taken into account.   The calculated visibility 
is convolved with a spectral resolution of 120, 
which approximately corresponds to the spectral resolution 
used for the Keck observations, 
and the Gaussian FWHM at each wavelength is derived 
for a projected baseline length of 10~m.  
Figure~\ref{omiCetUD_HKL} shows a comparison between 
the angular sizes predicted from our best-fit model for \mbox{$o$~Cet}\ from
1.5 to 4.0~\mbox{$\mu$m}\ and that observed by 
Ireland et al. (\cite{ireland04}).  The predicted angular sizes are 
in good agreement with the Keck observation, suggesting that 
the wavelength dependence of the angular size in the near-infrared 
largely reflects the opacity of \mbox{H$_2$O}\ lines.  
However, the Keck observation of \mbox{$o$~Cet}\ 
was carried out at phase 0.95, which is far from the phase 0.36 
of the TEXES observation of \mbox{$o$~Cet}.  Therefore, the apparent 
agreement between the observed and predicted near-infrared angular 
sizes of \mbox{$o$~Cet}\  should be regarded 
as preliminary, and this highlights again the importance of 
coordinated spectroscopic and interferometric observations.

\subsection{\mbox{R~Leo}}
\subsubsection{11~\mbox{$\mu$m}\ and near-infrared spectra}

\begin{figure}
\begin{center}
\resizebox{8.5cm}{!}{\rotatebox{-90}{\includegraphics{1207f8.ps}}}
\end{center}
\caption{Spectra of \mbox{R~Leo}\ in the 11~\mbox{$\mu$m}\ region at 
phase 0.75. 
The red dots represent the observed spectrum of \mbox{R~Leo}\ presented in 
WHT03b, while 
the green and blue solid lines represent the calculated spectra 
using the HITEMP database and the PS97 line list, respectively.  
The parameters of the best-fit model for \mbox{R~Leo}\ are 
\mbox{$T_{\rm hot}$}\ = 2000~K, 
\mbox{$R_{\rm hot}$}\ = 1.7~\mbox{$R_{\star}$}, \mbox{$N_{\rm hot}$}\  
= $1 \times 10^{21}$~\mbox{cm$^{-2}$}, \mbox{$T_{\rm cool}$}\ = 1200~K, 
\mbox{$R_{\rm cool}$}\ = 2.2~\mbox{$R_{\star}$}, 
and \mbox{$N_{\rm cool}$}\ = $7 \times 10^{21}$~\mbox{cm$^{-2}$}.  
The synthetic spectra are convolved with a Gaussian with a FWHM of 
0.013~\mbox{cm$^{-1}$}\ to account for the effects of the instrument as well 
as of the macro-turbulent velocity, and are redshifted by 
7~\mbox{km s$^{-1}$}\ to match the observation.  
The dashed lines represent the bandpasses used in the ISI 
observations. 
The positions of the \mbox{H$_2$O}\ lines whose 
intensity at 2000~K is stronger than 
$1 \times 10^{-24}$~cm molecule$^{-1}$ are marked with upper ticks 
(PS97 line list) and lower ticks (HITEMP database).  
These line positions are also redshifted by 7~\mbox{km s$^{-1}$}\ with 
respect to the rest wavenumber. 
}
\label{RLeoSp11mu}
\end{figure}

\begin{figure}
\begin{center}
\resizebox{8.5cm}{!}{\rotatebox{-90}{\includegraphics{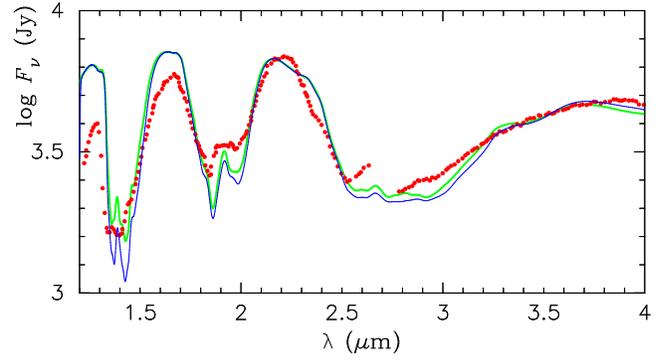}}}
\end{center}
\caption{Spectra of \mbox{R~Leo}\ in the region between 1.2 and 
4.0~\mbox{$\mu$m}. 
The red dots represent the KAO spectrum of \mbox{R~Leo}\ observed at 
phase 0.51 by Strecker et al. (\cite{strecker78}), while 
the green and blue solid lines represent the spectra from the 
best-fit model for phase 0.75, 
using the HITEMP database and the PS97 line list, respectively.  
The parameters of the model are given in the legend to
 Fig.~\ref{RLeoSp11mu}. 
The synthetic spectra are convolved to match the spectral resolution 
of the KAO observation ($R \simeq 50$).  
}
\label{RLeoSpKAO}
\end{figure}

For \mbox{R~Leo}, we find that the observed 11~\mbox{$\mu$m}\ spectrum and 
uniform disk diameters obtained by WHT03b can be best 
reproduced by a model with \mbox{$T_{\rm hot}$}\ = 2000~K, 
\mbox{$R_{\rm hot}$}\ = 1.7~\mbox{$R_{\star}$}, 
\mbox{$N_{\rm hot}$}\ = $1 \times 10^{21}$~\mbox{cm$^{-2}$}, 
\mbox{$T_{\rm cool}$}\ = 1200~K, \mbox{$R_{\rm cool}$}\ = 
2.2~\mbox{$R_{\star}$}, 
and \mbox{$N_{\rm cool}$}\ = $7 \times 10^{21}$~\mbox{cm$^{-2}$}.  
Figure~\ref{RLeoSp11mu} shows the synthetic spectra in the 11~\mbox{$\mu$m}\ 
region predicted by the best-fit model for \mbox{R~Leo}, together with 
the TEXES spectrum obtained at phase 0.75, 
which was read off Fig.~2 in WHT03b.  
The synthetic spectrum calculated with the HITEMP \mbox{H$_2$O}\ line list 
is represented with the green solid line, while that calculated with 
the PS97 \mbox{H$_2$O}\ line list is represented with the blue solid line.  
We estimated the flux contribution of the circumstellar dust shell 
in the 11~\mbox{$\mu$m}\ region based on the modeling of the spectral 
energy distribution (SED) and the visibility presented by Schuller et al. 
(\cite{schuller04}).  Their model SED for \mbox{R~Leo}, which can well reproduce 
the photometric observations in particular in the infrared, 
shows that the flux contribution of the dust shell at 11~\mbox{$\mu$m}\ is 
approximately 70\% of the total flux (see Fig.~3 of Schuller et al. 
\cite{schuller04}), and we adopted $f_{\rm dust} = 0.7$ in our 
calculation for \mbox{R~Leo}.  
The synthetic spectra are redshifted by 7~\mbox{km s$^{-1}$}\ to match the 
observation (see WHT03b). 
The figure shows that the synthetic spectra can reproduce the 
observed spectra without prominent spectral features in the bandpasses 
used for the ISI observations (marked with the dashed lines), 
which is a result of the filling-in 
effect due to the emission of the \mbox{H$_2$O}\ lines from the extended 
envelope.  The approximate depths and heights of 
the observed fine spectral features are also reproduced, as in 
the case of \mbox{$o$~Cet}\ discussed above.  

We also compare the predicted spectra in the near-infrared with that 
observed.  For \mbox{R~Leo}, the near-infrared spectrum obtained at phase 
0.51 by Strecker et al. (\cite{strecker78}) using the Kuiper Airborne 
Observatory (KAO) would be the spectrum closest to that expected at 
phase 0.75, at which the high-resolution 11~\mbox{$\mu$m}\  spectrum 
of \mbox{R~Leo}\ was obtained.  
Figure~\ref{RLeoSpKAO} shows a comparison between the 
spectrum of \mbox{R~Leo}\ observed by Strecker et al. (\cite{strecker78}) 
and those predicted from the best-fit model for \mbox{R~Leo}, which are 
convolved to match 
the spectral resolution of 50 used by 
Strecker et al. (\cite{strecker78}).  
As we will see below, the $K$-band visibility observed with the Keck 
telescope (Monnier et al. \cite{monnier04}) does not show 
a steep drop at low spatial frequencies characteristic of 
an object with an extended dust shell.  In addition, the SED modeling 
of Schuller et al. (\cite{schuller04}) shows that the flux contribution 
of the dust shell is $\sim$3\% at most in the near-infrared 
($\lambda \la 4$~\mbox{$\mu$m}).  Therefore, the flux contribution of the 
dust shell is neglected in the calculation of the synthetic spectra.
A glance of Fig.~\ref{RLeoSpKAO} 
reveals that the observed spectrum can be reasonably reproduced 
by the best-fit model for \mbox{R~Leo}\ for phase 0.75, and the difference 
between the observed and predicted spectra 
might be attributed to the mismatch in phase between the observation 
and the model.

\subsubsection{Angular diameters measured at 11~\mbox{$\mu$m}\ and in the 
near-infrared}

\begin{figure}
\begin{center}
\resizebox{8.5cm}{!}{\rotatebox{-90}{\includegraphics{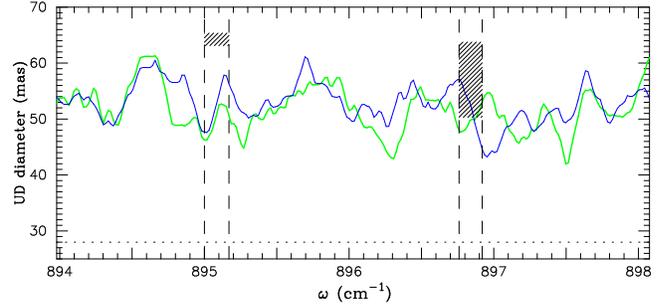}}}
\end{center}
\caption{
Uniform disk diameters calculated from the best-fit model for 
\mbox{R~Leo}\ in the 11~\mbox{$\mu$m}\ region for a projected baseline 
length of 30~m.  The green solid line represents 
the result calculated with the HITEMP database, while 
the blue solid line represents that calculated with the PS97 line 
list.  
The bandpasses used by WHT03b are marked 
with the dashed lines.  
The ranges of the diameters measured in these bandpasses are shown 
as the hatched regions.  
The dotted line represents the photospheric angular diameter 
adopted in the calculation. 
The calculated uniform disk diameters are redshifted by 
0.021~\mbox{cm$^{-1}$}\ with respect to the rest 
wavenumber, which corresponds to a radial velocity of 7~\mbox{km s$^{-1}$}, 
to match the observed spectrum shown in Fig.~\ref{RLeoSp11mu}. 
}
\label{RLeoVis11mu}
\end{figure}

\begin{figure}
\begin{center}
\resizebox{8.5cm}{!}{\rotatebox{-90}{\includegraphics{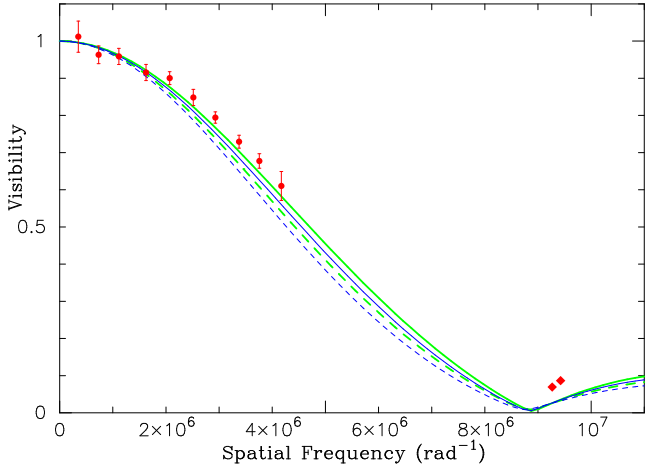}}}
\end{center}
\caption{Comparison between the $K$-band visibilities of \mbox{R~Leo}\ 
observed by Monnier et al. (\cite{monnier04}) and those predicted 
by the best-fit model for this object.  
The red filled circles represent the Keck observation, while the 
red filled diamonds represent the IOTA observation.  
The predicted visibilities which are convolved with a spectral 
resolution of 43 to match the Keck observation are plotted by 
the green solid line (calculated with the HITEMP database) and 
the blue solid line (calculated with the PS97 line list).  
The dashed lines represent the visibilities convolved 
with a spectral resolution of 7 to match the IOTA observation.  
The result calculated with the HITEMP database is plotted by the green 
dashed line, while that calculated with the PS97 line list is plotted 
by the blue dashed line.  
}
\label{RLeoVisK}
\end{figure}

We now compare the 11~\mbox{$\mu$m}\ angular diameters predicted from our
model for \mbox{R~Leo}\ 
with those obtained by the ISI observations.  Figure~\ref{RLeoVis11mu} 
shows a comparison between the observed uniform disk 
diameters and those predicted from the best-fit model for \mbox{R~Leo}.  
We tentatively adopt a photospheric angular diameter 
of 28~mas based on the measurements carried out at phase 0.24--0.28 by 
Perrin et al. (\cite{perrin99}) and 
examine the validity of this assumption by comparing with 
observed angular sizes in the near-infrared below.  
The uniform disk diameters are calculated from the visibilities 
excluding the dust shell for a projected baseline length of 30~m, 
as in the case of \mbox{$o$~Cet}.  
Figure~\ref{RLeoVis11mu} illustrates that the uniform disk 
diameters predicted within the bandpass at around 896.85~\mbox{cm$^{-1}$}\ 
are in agreement with those measured with ISI.   However, 
two measurements within this bandpass were carried out at different
epochs and phases, and therefore, it is necessary to compare the 
model predictions with the observations carried out at similar phases 
to that at the time of the TEXES observation of \mbox{R~Leo}.  

The ISI observations were carried out in 1999 October (phase 0.43) and 
in 2001 October and November (phase 0.77) for the bandpass at 
896.85~\mbox{cm$^{-1}$}\ and in 2001 November (phase 0.81) for the bandpass at 
895.1~\mbox{cm$^{-1}$}.  Since our model for \mbox{R~Leo}\ is based on the 
fitting to the TEXES spectrum obtained in 2000 December at phase 0.75, 
the predicted uniform disk diameters 
should be close to the values observed at similar phases, that is, 
$62.62 \pm 1.14$~mas at 896.85~\mbox{cm$^{-1}$}\ at phase 0.77 and 
$64.24 \pm 1.12$~mas at 895.1~\mbox{cm$^{-1}$}\ at phase 0.81.  
Although the predicted uniform disk 
diameters in the two bandpasses are somewhat smaller than these 
observed values obtained at phases close to 0.75, 
this difference may be attributed to a cycle-to-cycle variation 
and/or deviation from circular symmetry as WHT03a discuss for 
\mbox{$o$~Cet}.  Given such a possible temporal variation of the 
angular size and a presence of asymmetries, we can conclude 
that our model for the warm water vapor envelope can fairly 
reproduce the observed 11~\mbox{$\mu$m}\ uniform disk diameter 
of \mbox{R~Leo}. 

As in the case of \mbox{$o$~Cet}, we compare near-infrared visibilities 
predicted by the best-fit model for \mbox{R~Leo}\ with those observed 
at phases close to 0.75.  Monnier et al. (\cite{monnier04}) have 
recently combined interferometric data from Keck aperture masking 
and the Infrared Optical Telescope Array (IOTA) for ten objects 
including \mbox{R~Leo}.  They observed \mbox{R~Leo}\ at phase 0.7, 
very close to the phase 0.75.  Figure~\ref{RLeoVisK} shows 
a comparison between the $K$-band visibilities obtained by 
Monnier et al. (\cite{monnier04}) and those predicted from the 
best-fit model for \mbox{R~Leo}.  The observed visibility at low spatial 
frequencies ($\la 4 \times 10^6$~rad$^{-1}$) does not show a sharp drop, 
which would be present if there were flux contribution from the 
dust shell.  Therefore, we include no dust emission in the calculation 
of the $K$-band visibilities.  The predicted visibilities are 
spectrally convolved with a Gaussian with a FWHM of 0.053~\mbox{$\mu$m}\ 
($R \simeq 43$) to match the width of the narrow-band filter used 
for the Keck observation.  
For comparison with the IOTA data, which were obtained 
with a broadband \mbox{$K^{\prime}$}\ filter, the predicted visibilities are 
convolved with a top-hat function centered at 2.16~\mbox{$\mu$m}\ 
with $\Delta \lambda = 0.32$~\mbox{$\mu$m}\ ($R \simeq 7$).  
Figure~\ref{RLeoVisK} demonstrates that the visibility obtained by 
Keck aperture masking (red filled circles) can be well reproduced 
by the best-fit model, 
while the predicted visibilities are too low compared to 
the visibility points obtained with IOTA (red filled diamonds).  
This discrepancy at high spatial frequencies may be attributed 
to the simplicity of our model and/or a cycle-to-cycle variation 
which might have taken place between the Keck+IOTA observations 
(2000 January and February) and the TEXES observation 
(2000 December).

\begin{figure}
\begin{center}
\resizebox{8.5cm}{!}{\rotatebox{-90}{\includegraphics{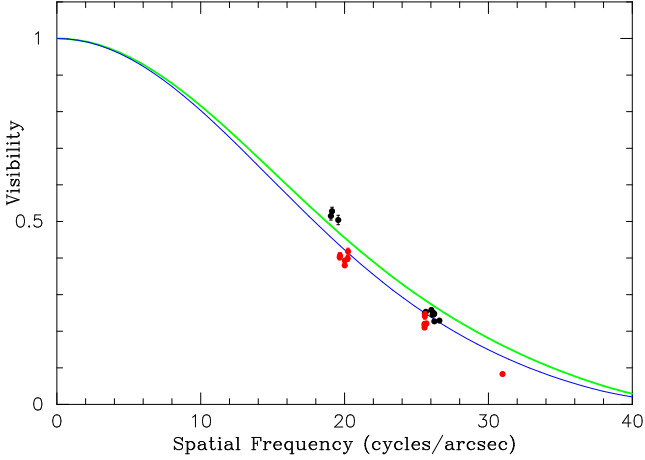}}}
\end{center}
\caption{Comparison between the \mbox{$L^{\prime}$} -band visibilities 
of \mbox{R~Leo}\ 
observed by Chagnon et al. (\cite{chagnon02}) and those predicted 
by the best-fit model for this object.  
The black filled circles represent the visibilities observed in 
2000 March (phase 0.81--0.82), while the red filled circles represent 
those observed in 2000 November (phase 0.61--0.64).  
The green solid line represents the predicted visibility calculated 
with the HITEMP database, while the blue solid line represents that 
calculated with the PS97 line list.  
}
\label{RLeoVisL}
\end{figure}

Chagnon et al. (\cite{chagnon02}) measured \mbox{$L^{\prime}$}-band visibility 
of \mbox{R~Leo}\ at two epochs, phase $\sim$0.6 and $\sim$0.8, both of which 
are rather close to the phase 0.75.  Figure~\ref{RLeoVisL} shows a 
comparison between the observed \mbox{$L^{\prime}$}-band visibilities 
of \mbox{R~Leo}\ and those predicted by the best-fit model for this object.  
Since the flux contribution of the dust shell is expected to be
$\sim$3\% at most in the \mbox{$L^{\prime}$}\ band as mentioned above, 
no flux contribution from the dust shell is included in the calculation 
of the \mbox{$L^{\prime}$}-band visibilities.  
The predicted visibilities are spectrally convolved with a top-hat 
function centered at 3.8~\mbox{$\mu$m}\ with 
$\Delta \lambda$ = 0.54~\mbox{$\mu$m}\ 
to match the observed data.  The figure illustrates that the observed 
visibilities can be fairly reproduced by our model, 
given the slight mismatch in phase as well as a possible cycle-to-cycle 
variation of the warm water vapor envelope.  
This agreement of the \mbox{$L^{\prime}$}-band visibility lends further 
support to the photospheric angular diameter that we adopted and 
our basic picture of the warm water vapor envelope.

\begin{figure}
\begin{center}
\resizebox{8.5cm}{!}{\rotatebox{-90}{\includegraphics{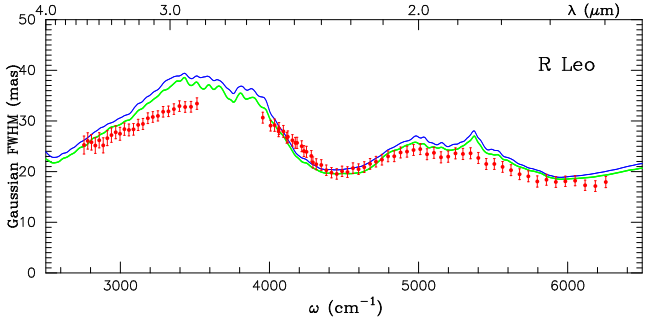}}}
\end{center}
\caption{Wavelength dependence of the angular size (Gaussian FWHM) 
predicted from the best-fit model for \mbox{R~Leo}\ for a projected baseline 
length of 10~m.  The red dots represent the angular size of \mbox{R~Leo}\ 
observed by Ireland et al. (\cite{ireland04}).  
The green solid line represents the result 
obtained with the HITEMP database, while the blue solid line represents 
the result with the PS97 line list.  No flux contribution from the dust 
shell is included in the calculation shown.  
}
\label{RLeoUD_HKL}
\end{figure}

We also compare the wavelength dependence of the angular size predicted 
in the near-infrared with 
that observed by Ireland et al. (\cite{ireland04}).  Their observation 
of \mbox{R~Leo}\ in the wavelength region between 1.2 and 
3.6~\mbox{$\mu$m}\ with the Keck telescope was carried 
out at phase 0.74, which is very close to the phase of 0.75 at which 
the high-resolution 11~\mbox{$\mu$m}\  spectrum was obtained.  
We calculate the near-infrared angular size using the best-fit model 
for \mbox{R~Leo}\  in the same manner as in the case of \mbox{$o$~Cet}.  
Comparison between the observed Gaussian FWHM of \mbox{R~Leo}\ and those
predicted from our model shown in Fig.~\ref{RLeoUD_HKL} 
illustrates that our model can reproduce the observed 
Gaussian FWHM from 1.5 to 3.6~\mbox{$\mu$m}\ quite well.  
This agreement of the near-infrared angular size of \mbox{R~Leo}\ 
suggests again that the \mbox{H$_2$O}\ lines govern the 
wavelength dependency of the angular size of Mira variables in this 
wavelength regime.  

\subsection{\mbox{$\chi$~Cyg}}

\subsubsection{11~\mbox{$\mu$m}\ and near-infrared spectra}

\begin{figure}
\begin{center}
\resizebox{8.5cm}{!}{\rotatebox{-90}{\includegraphics{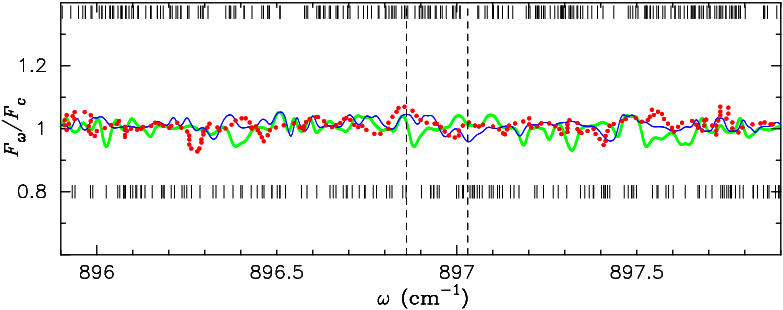}}}
\end{center}
\caption{Spectra of \mbox{$\chi$~Cyg}\ in the 11~\mbox{$\mu$m}\ region 
at phase 0.4. 
The red dots represent the observed spectrum of \mbox{$\chi$~Cyg}\ 
presented in WHT03b, while 
the green and blue solid lines represent the calculated spectra 
using the HITEMP database and the PS97 line list, respectively.  
The parameters of the best-fit model for \mbox{$\chi$~Cyg}\ are 
\mbox{$T_{\rm hot}$}\ = 2000~K, \mbox{$R_{\rm hot}$}\ =
 1.5~\mbox{$R_{\star}$}, \mbox{$N_{\rm hot}$}\  
= $1 \times 10^{21}$~\mbox{cm$^{-2}$}, \mbox{$T_{\rm cool}$}\ = 1200~K,
 \mbox{$R_{\rm cool}$}\ = 2.2~\mbox{$R_{\star}$}, 
and \mbox{$N_{\rm cool}$}\ = $1 \times 10^{21}$~\mbox{cm$^{-2}$}.  
The synthetic spectra are convolved with a Gaussian with a FWHM of 
0.013~\mbox{cm$^{-1}$}\ to account for the effects of the instrument as well 
as of the macro-turbulent velocity, and are blueshifted by 
22~\mbox{km s$^{-1}$}\ to match the observation.  
The dashed lines represent the bandpasses used in the ISI 
observations. 
The positions of the \mbox{H$_2$O}\ lines whose 
intensity at 2000~K is stronger than 
$1 \times 10^{-24}$~cm molecule$^{-1}$ are marked with upper ticks 
(PS97 line list) and lower ticks (HITEMP database).  
These line positions are also blueshifted by 22~\mbox{km s$^{-1}$}\ with 
respect to the rest wavenumber. 
}
\label{chiCygSp11mu}
\end{figure}

For \mbox{$\chi$~Cyg}, 
we find that the observed 11~\mbox{$\mu$m}\ spectrum and the 
uniform disk diameter obtained by WHT03b can be best 
reproduced by a model with \mbox{$T_{\rm hot}$}\ = 2000~K, 
\mbox{$R_{\rm hot}$}\ = 1.5~\mbox{$R_{\star}$}, 
\mbox{$N_{\rm hot}$}\ = $1 \times 10^{21}$~\mbox{cm$^{-2}$}, 
\mbox{$T_{\rm cool}$}\ = 1200~K, 
\mbox{$R_{\rm cool}$}\ = 2.2~\mbox{$R_{\star}$}, 
and \mbox{$N_{\rm cool}$}\ = $1 \times 10^{21}$~\mbox{cm$^{-2}$}.  
Figure~\ref{chiCygSp11mu} shows a comparison between the TEXES 
spectrum of \mbox{$\chi$~Cyg}, which was obtained in 2001 mid-June at 
phase 0.4, and those predicted by the above best-fit model for this object.  
The flux contribution of the circumstellar dust shell at 
11~\mbox{$\mu$m}\ was estimated from the visibility function 
at 11~\mbox{$\mu$m}\ obtained by Danchi et al. (\cite{danchi94}), which 
indicates a steep drop at low spatial frequencies resulting from 
the presence of the extended dust shell.  The amount of the visibility drop, 
which is equal to the fraction of the flux contribution of the 
dust shell, is approximately 0.6, and we adopted $f_{\rm dust} = 0.6$ 
in our calculation for \mbox{$\chi$~Cyg}.  The synthetic spectra are 
blueshifted by 22~\mbox{km s$^{-1}$}\ to match the observation (see WHT03b).  
Figure~\ref{chiCygSp11mu} demonstrates that the observed spectrum 
(red dots) can be well reproduced by the model.  Like in the cases 
of \mbox{$o$~Cet}\ and \mbox{R~Leo}\ discussed above, the filling-in 
effect due to \mbox{H$_2$O}\ line emission from the extended envelope 
leads to the featureless, continuum-like spectra.  

\begin{figure}
\begin{center}
\resizebox{8.5cm}{!}{\rotatebox{-90}{\includegraphics{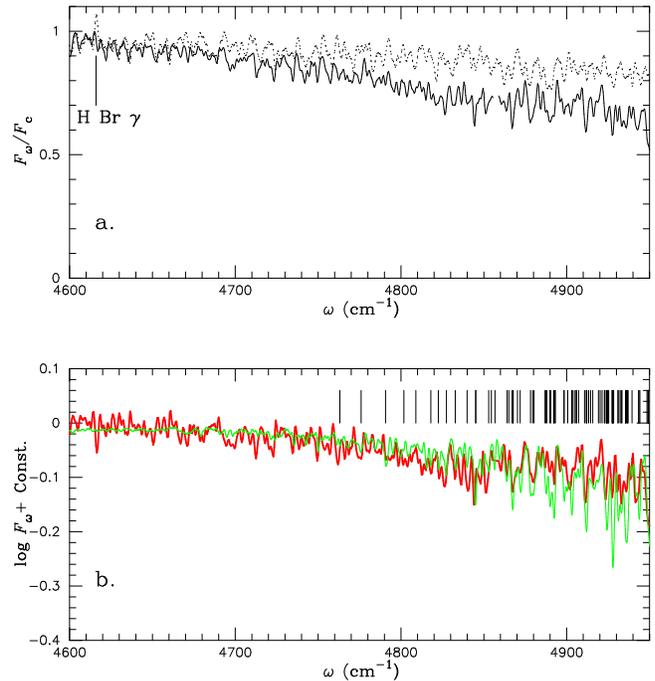}}}
\end{center}
\caption{Spectra of \mbox{$\chi$~Cyg}\ in the region between 4600 and 
4950~\mbox{cm$^{-1}$}. 
{\bf a}: The dotted line represents the spectrum of \mbox{$\chi$~Cyg}\ 
obtained at phase 0.00, while the solid line represents the spectrum at 
phase 0.42 (Wallace \& Hinkle \cite{wallace97}).  
{\bf b}: The thick red solid line represents the ratio between the spectra 
observed at phase 0.00 and 0.42.  The thin green solid line 
represents the spectrum predicted by the best-fit model for \mbox{$\chi$~Cyg}, 
whose parameters are given in the legend to 
Fig.~\ref{chiCygSp11mu}. 
The synthetic spectrum is convolved to match the spectral resolution 
of $R \simeq 3000$ used in the observations by 
Wallace \& Hinkle (\cite{wallace97}).  
}
\label{chiCygSpK}
\end{figure}

The presence of such a dense water vapor envelope around \mbox{$\chi$~Cyg}, 
though the \mbox{H$_2$O}\ column density is smaller than that in 
\mbox{$o$~Cet}\ and \mbox{R~Leo}, 
may appear to be in conflict with the classification of this 
object as an S star.  S stars have C/O ratios close to unity, 
and therefore, neither oxygen-bearing nor carbon-bearing molecules 
are abundant in the atmosphere, with 
most of carbon and oxygen atoms locked up in CO molecules.  
In fact, the \mbox{H$_2$O}\ lines observed for \mbox{$\chi$~Cyg}\ in 
the 4~\mbox{$\mu$m}\ region are significantly weaker than those
observed in oxygen-rich Miras (Lebzelter et al. \cite{lebzelter01}).  
However, 
there are some pieces of observational evidence for the presence of 
a significant amount of water vapor in \mbox{$\chi$~Cyg}\ near minimum light.  
Based on the \mbox{H$_2$O}\ absorption features in the 9000~\AA\ region, 
Spinrad \& Vardya (\cite{spinrad66b}) estimated the \mbox{H$_2$O}\
column density in \mbox{$\chi$~Cyg}\ to be 
$3 \times 10^{21}$~\mbox{cm$^{-2}$}\ at 
minimum light, which is in agreement with the result we obtained 
above.  Wallace \& Hinkle (\cite{wallace97}) present a series of $K$-band 
spectra of \mbox{$\chi$~Cyg}\ observed at various phases 
with a spectral resolution of $\sim$3000 using a Fourier Transform 
Spectrometer (FTS), and the 
spectra exhibit an increasing absorption in the region between 
$\sim$4700~\mbox{cm$^{-1}$}\ and 4950~\mbox{cm$^{-1}$}\ from 
maximum light (phase 0.00) to minimum light 
(phase 0.42--0.48).  Figure~\ref{chiCygSpK}a shows the spectra 
acquired at phase 0.00 and 0.42\footnote{Available at 
ftp://ftp.noao.edu/catalogs/medresIR/K\_band/}, and the thick red solid  
line in Fig.~\ref{chiCygSpK}b represents the ratio of these two spectra 
(i.e., spectrum at 0.42 divided by spectrum at 0.00).  
This ratioed spectrum 
clearly shows the enhanced absorption which appears between 
$\sim$4700 and 4950~\mbox{cm$^{-1}$}\ near minimum light.  
The thin green solid line in Fig.~\ref{chiCygSpK}b represents the 
\mbox{H$_2$O}\ spectrum predicted by the above best-fit model for 
\mbox{$\chi$~Cyg}\ using the HITEMP database.  
Dyck et al. (\cite{dyck84}) estimated that more than 90\% of the 
observed flux between 2.2 and 4.8~\mbox{$\mu$m}\ comes from the star based 
on the modeling of Rowan-Robinson \& Harris (\cite{rowanrobinson83}). 
Therefore, the flux contribution of the dust shell is not included 
in the calculation of the synthetic spectrum.  
As the figure shows, the absorption observed between 
$\sim$4700 and 4950~\mbox{cm$^{-1}$}\ is reproduced by the synthetic spectrum, 
and the individual absorption features seen in the ratioed 
spectrum can be identified as \mbox{H$_2$O}\ lines, 
whose line positions are marked with the ticks in the figure.  

\begin{figure}
\begin{center}
\resizebox{8.5cm}{!}{\rotatebox{-90}{\includegraphics{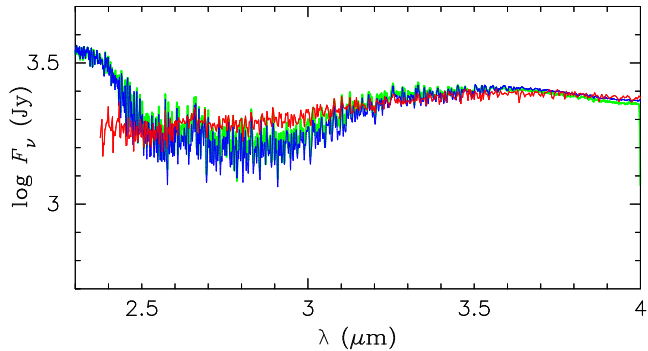}}}
\end{center}
\caption{Spectra of \mbox{$\chi$~Cyg}\ in the region between 2.3 and 
4.0~\mbox{$\mu$m}. 
The red solid line represents the spectrum of \mbox{$\chi$~Cyg}\ 
obtained at phase 0.5 
with ISO/SWS (Vandenbussche et al. \cite{vandenbussche02}), while 
the green and blue solid lines represent the spectra from the 
best-fit model for phase 0.4, 
using the HITEMP database and the PS97 line list, respectively.  
The parameters of the model are given in the legend to 
Fig.~\ref{chiCygSp11mu}. 
The synthetic spectra are convolved to match the spectral resolution 
of the ISO/SWS observation ($R \simeq 1500$), and no flux contribution 
from the dust shell is included.  
}
\label{chiCygSpISO}
\end{figure}

Finally, more definitive evidence for the presence of a large amount of 
water vapor in \mbox{$\chi$~Cyg}\ near minimum light can be found in 
the near-infrared spectrum obtained with ISO/SWS.  
Vandenbussche et al. (\cite{vandenbussche02}) present 
the ISO/SWS spectrum of \mbox{$\chi$~Cyg}\ obtained on 1998 May 10 
(UT 07:39:48) at phase 0.5 in the wavelength region between 2.3 and
4.1~\mbox{$\mu$m}.  
We retrieved this ISO/SWS spectrum of \mbox{$\chi$~Cyg}\ from the ISO 
data archive and plot it in Fig.~\ref{chiCygSpISO}.   
The spectrum clearly exhibits the absorption feature centered at 
2.7~\mbox{$\mu$m}\ due to the \mbox{H$_2$O}\ 
$\nu_1$ and $\nu_3$ fundamental bands.  
Also plotted in the figure are the spectra predicted from the above 
best-fit model for \mbox{$\chi$~Cyg}, and the figure demonstrates that 
the predicted 
spectra are in fair agreement with that observed, given the simplicity 
of our model as well as a slight mismatch in phase between the TEXES 
observation and the ISO/SWS observation.  

The above observational 
results show that water vapor is present in the S-type Mira variable 
\mbox{$\chi$~Cyg}\ near minimum light, and that the amount of water vapor is 
comparable to those found in the oxygen-rich Miras \mbox{$o$~Cet}\ and 
\mbox{R~Leo}.  In general, 
Mira variables exhibit the strongest molecular absorption near 
minimum light (e.g., Lan\c{c}on \& Wood \cite{lancon00}), which implies 
that the temperature may be the lowest at minimum light and that 
the formation of water vapor may be significantly promoted at such 
low temperatures.  However, detailed calculations using 
dynamical atmospheres are necessary for understanding 
the formation of \mbox{H$_2$O}\ in Mira variables with C/O ratios very close 
to unity.  

\subsubsection{Angular diameters measured at 11~\mbox{$\mu$m}\ and in the 
near-infrared}

\begin{figure}
\begin{center}
\resizebox{8.5cm}{!}{\rotatebox{-90}{\includegraphics{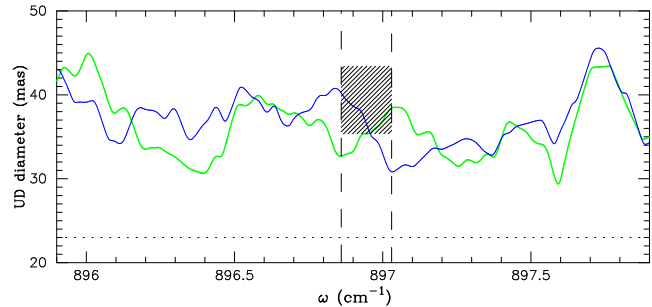}}}
\end{center}
\caption{
Uniform disk diameters calculated from the best-fit model for 
\mbox{$\chi$~Cyg}\ in the 11~\mbox{$\mu$m}\ region for a projected baseline 
length of 30~m. 
The green solid line represents the result 
obtained with the HITEMP database, while the blue solid line represents 
the result with the PS97 line list.
The bandpass used by WHT03b is marked with the dashed lines.  
The range of the diameters measured in this bandpass is shown 
as the hatched region.  
The dotted line represents the photospheric angular diameter 
adopted in the calculation. 
The calculated uniform disk diameters are blueshifted by 
0.0658~\mbox{cm$^{-1}$}\ with respect to the rest 
wavenumber, which corresponds to a radial velocity of 
$-22$~\mbox{km s$^{-1}$}, 
to match the observed spectrum shown in Fig.~\ref{chiCygSp11mu}. 
}
\label{chiCygVis11mu}
\end{figure}

In Fig.~\ref{chiCygVis11mu}, 
we compare the angular diameter of \mbox{$\chi$~Cyg}\ measured 
in the 11~\mbox{$\mu$m}\ 
region and those predicted by the above best-fit model for this object.
The phase of \mbox{$\chi$~Cyg}\ at the time of the ISI observation is 0.51 
(2001 July and August, see WHT03b), which is quite close to the phase 
0.4 at which the TEXES spectrum was obtained.  
We adopt a photospheric angular diameter of 23~mas based on a 
$K^{\prime}$-band uniform disk diameter measured by 
Mennesson et al. (\cite{mennesson02}) at phase 0.38, which is also 
close to the phase 0.4.  As in the cases of \mbox{$o$~Cet}\ and 
\mbox{R~Leo}, we check the validity of the assumed photospheric angular 
diameter by calculating the near-infrared angular diameters below.  
The uniform disk diameters are calculated from the visibilities
excluding the dust shell for a projected baseline length of 30~m.  
Figure~\ref{chiCygVis11mu} shows that the uniform disk diameters predicted 
in the ISI bandpass are in agreement with the observed value of 
$39.38 \pm 4.02$~mas.  

We also compare the angular diameters predicted in the near-infrared 
with those observed.  Mennesson et al. (\cite{mennesson02}) 
observed \mbox{$\chi$~Cyg}\ in the \mbox{$K^{\prime}$}\ and 
\mbox{$L^{\prime}$}\ bands and derived 
uniform disk diameters of $23.24 \pm 0.08$~mas and 
$30.40^{+3.30}_{-7.28}$~mas, respectively.  
We calculate the uniform disk diameters in these bands, with the 
\mbox{$K^{\prime}$}- and \mbox{$L^{\prime}$}-band filters represented 
with top-hat functions 
centered at 2.16~\mbox{$\mu$m}\ with $\Delta \lambda$ = 0.32~\mbox{$\mu$m}\ 
and at 3.8~\mbox{$\mu$m}\ with $\Delta \lambda$ = 0.54~\mbox{$\mu$m}, 
respectively.  
The \mbox{$K^{\prime}$}-band uniform disk diameters calculated with the HITEMP 
database and the PS97 line list for a projected baseline length of 20~m 
are 24.4~mas and 24.7~mas, respectively.  
These predicted \mbox{$K^{\prime}$}-band uniform disk diameters are 
in reasonable 
agreement with the observed value, which was measured at phase 0.38, 
very close to the phase 0.4 of the TEXES observation.  
The \mbox{$L^{\prime}$}-band uniform disk diameters calculated with the HITEMP 
database and the PS97 line list are 27.1~mas and 28.1~mas, respectively.  
While these predicted \mbox{$L^{\prime}$}-band uniform disk diameters 
are also in 
agreement with the observed one within the error of the measurement, 
it should be noted that the \mbox{$L^{\prime}$}-band observations of 
Mennesson et al. (\cite{mennesson02}) were carried out at phase 0.81, 
which makes it difficult to compare the observed 
angular diameter with that predicted by the best-fit model based on 
the TEXES spectrum obtained at phase 0.4.  Therefore, we can conclude 
that the near-infrared angular diameters predicted by our model for 
\mbox{$\chi$~Cyg}\ are in rough agreement with those observed, 
while more rigorous tests are necessary using contemporaneous 
interferometric observations in the near-infrared and in the 
mid-infrared.  

\section{Discussion}

Our two-layer model for the warm water vapor envelope 
around three Mira variables can reproduce the observed increase of the 
angular diameter from the near-infrared to the 11~\mbox{$\mu$m}\ region 
and, simultaneously, the observed high-resolution 11~\mbox{$\mu$m}\ spectra 
which appear to be featureless.  It is obvious, 
however, that our ad hoc model is not a unique solution, and that 
physical processes responsible for the formation of such warm water 
vapor layers also remain to be theoretically understood.  

Intensity profiles predicted by dynamical 
model atmospheres for Mira variables exhibit structures consisting of 
two components.  Tej et al. (\cite{tej03}) show that the formation of 
\mbox{H$_2$O}\ layers behind shock fronts leads to intensity profiles 
consisting of a step-like structure close to the star and/or 
a tenuous halo-like structure extending to a few stellar radii.  
The intensity profiles predicted by our models for the three Mira
variables studied here also show such a step-like structure and 
a more extended component, as can be seen in Fig.~\ref{omiCetVis11mu}c.  
This implies the 
possibility that the large-amplitude pulsation in Mira variables 
is responsible for the warm molecular envelope in these objects, and 
the next logical step would be to calculate synthetic spectra as well 
as intensity profiles using dynamical model atmospheres such as 
presented in Tej et al. (\cite{tej03}) and 
extensive \mbox{H$_2$O}\ line lists. 

Hinkle \& Barnes (\cite{hinkle79}) analyzed \mbox{H$_2$O}\ lines in 
high-resolution near-infrared spectra (4000--6700~\mbox{cm$^{-1}$}) 
of \mbox{R~Leo}\ 
and identified two distinct \mbox{H$_2$O}\ layers, that is, a warm 
component with a temperature of $\sim$1700~K and a cool component with a
temperature of $\sim$1100~K.  
This result seems to be in agreement with the 
temperatures of the \mbox{H$_2$O}\ layers we obtained for 
\mbox{R~Leo}\ in the present work.  
The \mbox{H$_2$O}\ column densities of the cool component derived by 
Hinkle \& Barnes (\cite{hinkle79}) range from 
$10^{21}$--$10^{22}$~\mbox{cm$^{-2}$}, and these values also agree with the 
$7 \times 10^{21}$~\mbox{cm$^{-2}$}\ we derived for the cool 
\mbox{H$_2$O}\  layer.  
On the other hand, the \mbox{H$_2$O}\ column densities of the warm 
component they derived are $10^{19}$--$10^{20}$~\mbox{cm$^{-2}$}, 
which are significantly smaller 
than the $1 \times 10^{21}$~\mbox{cm$^{-2}$}\ derived here.  
Hinkle \& Barnes (\cite{hinkle79}) also note that the \mbox{H$_2$O}\ lines 
originating in the warm component are never predominant compared 
to those originating in the cool component, which shows a marked 
contrast to our result that the hot \mbox{H$_2$O}\ layer contributes 
significantly in the spectrum as well as in the intensity profile. 
However, Hinkle \& Barnes (\cite{hinkle79}) derived the \mbox{H$_2$O}\ 
column densities at phases 0.94, 0.00, 0.14, and 0.20, but not at phases 
between 0.5 and 0.9 because of the blend of \mbox{H$_2$O}\ lines originating 
in the two components.  Since our model for \mbox{R~Leo}\ is based on the 
TEXES spectrum obtained at phase 0.75, the above discrepancy of the 
\mbox{H$_2$O}\ column density of the warm component might be due to the 
mismatch in phase.  Comparison with synthetic spectra using dynamical 
model atmospheres would be necessary to derive physical parameters such 
as excitation temperature and column density from the 
high-resolution spectra observed near minimum light which are severely
affected by the blend of lines.

\section{Concluding remarks}

We have shown that our two-layer model for the warm water vapor envelope 
around the Mira variables \mbox{$o$~Cet}, \mbox{R~Leo}, and \mbox{$\chi$~Cyg}\ 
can reproduce the observed increase of the 
angular diameter from the near-infrared to the 11~\mbox{$\mu$m}\ region 
and, simultaneously, the observed high-resolution 11~\mbox{$\mu$m}\  spectra. 
While a number of \mbox{H$_2$O}\ pure rotation lines are present in the 
wavelength region observed with ISI and strong absorption can be 
expected, the absorption lines are filled in by the emission lines 
originating in the extended part of the water vapor envelope, which 
leads to the featureless, continuum-like spectra.  This filling-in 
effect masks the spectroscopic fingerprints of the warm water vapor 
envelope, but its presence manifests itself as an increase of the 
angular diameter 
from the near-infrared to the mid-infrared: invisible for spectrometers 
but not for interferometers.  
The radii, temperatures, and \mbox{H$_2$O}\ column densities of the hot 
\mbox{H$_2$O}\ layer in the three Mira variables studied here 
are derived to be 1.5--1.7~\mbox{$R_{\star}$}, 1800--2000~K, and 
(1--$5) \times 10^{21}$~\mbox{cm$^{-2}$}, respectively.  
The cool \mbox{H$_2$O}\ 
layer is found to have temperatures of 1200--1400~K, extending to 
2.2--2.5~\mbox{$R_{\star}$}\ with \mbox{H$_2$O}\ column densities of 
(1--$7) \times 10^{21}$~\mbox{cm$^{-2}$}.  
Our models which reproduce the spectra and angular diameters observed 
at 11~\mbox{$\mu$m}\ have turned 
out to also reproduce the spectra and the visibilities as well as the 
angular diameters observed in the 
near-infrared.  Comparison between the near-infrared angular sizes 
predicted for \mbox{$o$~Cet}\ as well as for \mbox{R~Leo}\ and those 
observed with the Keck telescope 
suggests that the wavelength dependence of the angular size of Mira 
variables in the near-infrared largely reflects the \mbox{H$_2$O}\  opacity.  

We have also found evidence for the presence of a large amount of 
water vapor in \mbox{$\chi$~Cyg}\ near minimum light in the FTS spectra 
obtained in the $K$ band as well as in the ISO/SWS spectrum between 2.3 
and 4.1~\mbox{$\mu$m}.  The observed \mbox{H$_2$O}\ 
absorption features can be reasonably well reproduced by our 
water vapor envelope model whose parameters are derived from the 
comparison of the spectrum and the angular diameter observed 
at 11~\mbox{$\mu$m}.  While S-type Miras are known to exhibit significantly 
weak spectral features due to oxygen-bearing molecules compared 
to oxygen-rich Miras, water vapor of an amount comparable to that 
in oxygen-rich Miras is present in \mbox{$\chi$~Cyg}\ near minimum light.  

In order to test our model for the warm water vapor envelope further, 
contemporaneous spectroscopic and interferometric observations are 
indispensable, given the variations of the spectra and the 
angular size not only with phase but also on a time scale longer 
than the variability period.  ISI observations using many more 
bandpasses in the 11~\mbox{$\mu$m}\ region and/or with even higher spectral 
resolution would also be useful for this purpose.

\begin{acknowledgement}
The author would like to thank Dr.~M.~Ireland for kindly providing 
the result of the Keck observations of \mbox{$o$~Cet}\ and \mbox{R~Leo}\ in 
electronic format, 
and Dr.~J.~Weiner for the information on the dates of the TEXES 
observations of \mbox{R~Leo}\ and \mbox{$\chi$~Cyg}.  
The author is also indebted to Dr.~T.~Driebe and the referee 
Dr.~D.~D.~S.~Hale for valuable and constructive comments.  
\end{acknowledgement}

\end{document}